\documentclass[11pt]{cernrep}
\usepackage{graphicx,epsfig}
\usepackage{cite}
\newcommand{\leqsim}{\,\raisebox{-0.6ex}{$\buildrel < \over \sim$}\,}
\newcommand{\geqsim}{\,\raisebox{-0.6ex}{$\buildrel > \over \sim$}\,}
\begin{document}

\title{Low-$x$ physics and $W$ and $Z$ production at the LHC
}

\author{A. M. Cooper-Sarkar$^1$}
\institute{$^1$Particle Physics, Oxford University, Denys Wilkinson Building, Keble Rd, OXFORD, OX1 3RH, UK
}

\maketitle


\section{INTRODUCTION}
\label{sec:lowx;amcs_intro}

\begin{figure}[tbp]
\centerline{
\epsfig{figure=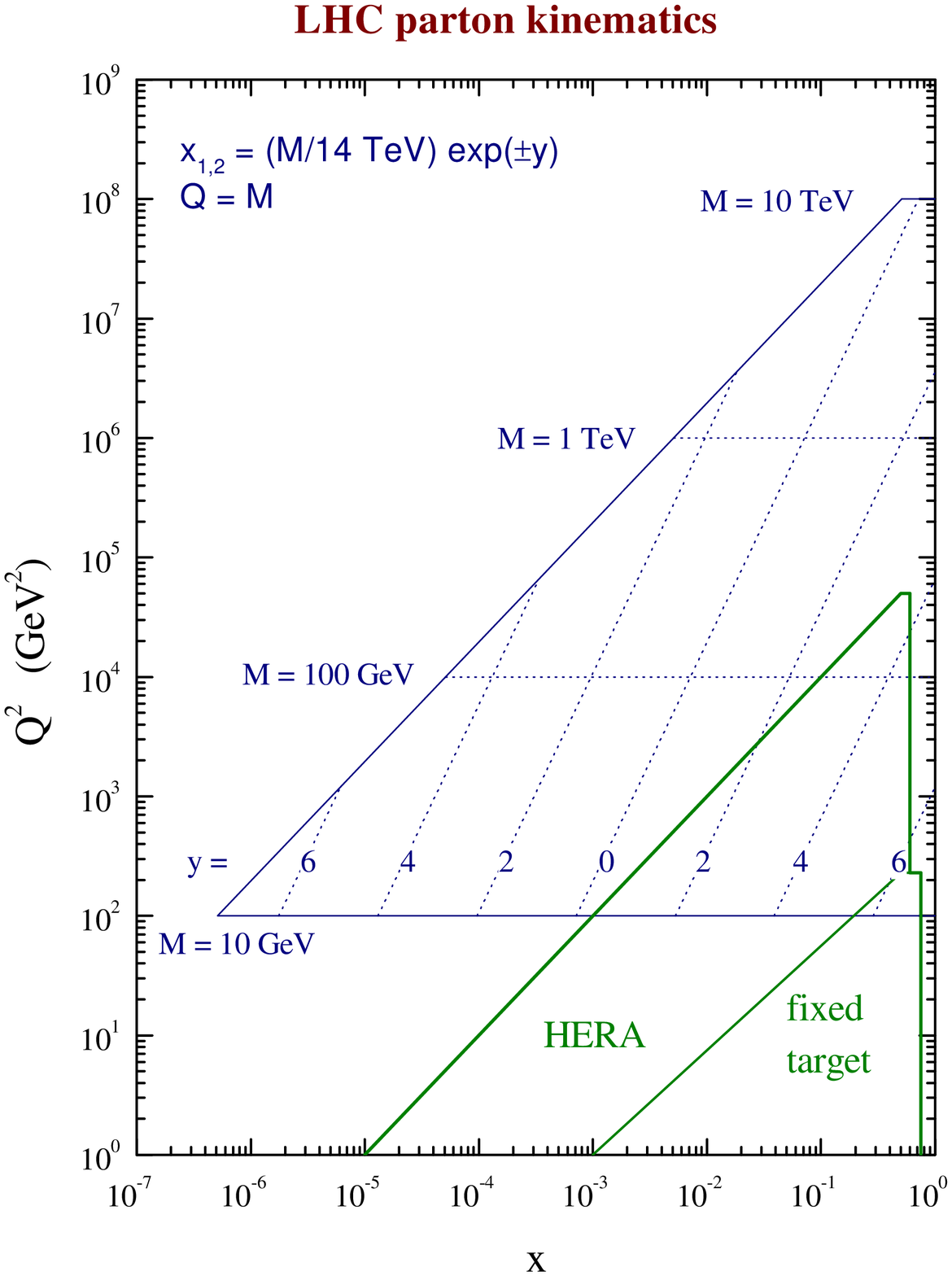,width=0.5\textwidth}
\epsfig{figure=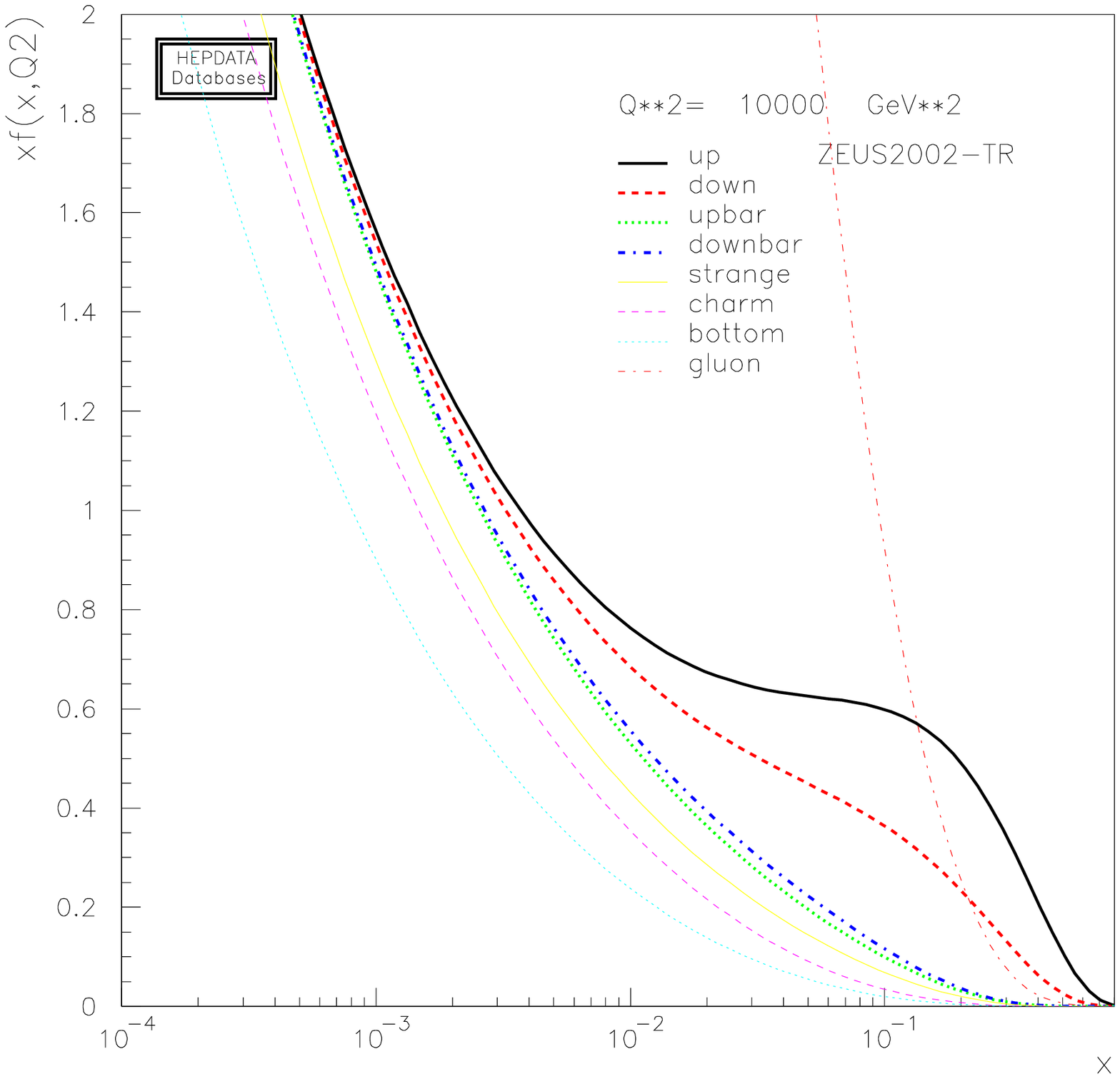,width=0.5\textwidth}}
\caption {Left plot: The LHC kinematic plane (thanks to James Stirling).
Right plot: PDF distributions at $Q^2 = 10,000$~GeV$^2$.}
\label{fig:kin/pdfs}
\end{figure}

The kinematic plane for LHC is shown in Fig.~\ref{fig:kin/pdfs}, which 
translates the kinematics for producing a state of mass $M$ and rapidity $y$ 
into the deep inelastic scattering variables, $Q^2$, the scale of the 
hard sub-process, and the Bjorken $x$ 
values of the participating partons. The scale of the process is given by
$Q^2 = M^2$ and the Bjorken $x$ values by, $x_1 = (M/ \surd{s}) exp(y)$, and, 
$x_2 = (M/ \surd{s}) exp(-y)$, where $y$ is the parton rapidity,  
$y = \frac{1}{2} ln{\frac{(E+pl)}{(E-pl)}}$. Thus, at central rapidity, 
these $x$ 
values are equal, but as one moves away from central rapidity, one parton movesto higher $x$ and one to lower $x$, as illustrated by the lines of constant $y$ 
on the plot.
The first physics to be studied at the LHC will be at relatively 
low scales, where the large cross-sections ensure that even low 
luminosity running will yield copious numbers of events. Thus the LHC will 
begin by studying standard model (SM) physics, calibrating our knowledge of 
the detectors on these well known processes. Study of 
Fig.~\ref{fig:kin/pdfs} makes it clear that the cross-sections for these 
processes are only well known if 
the parton distribution functions (PDFs) of the proton are 
well known at small-$x$. This assumes that the theoretical 
formalism of NLOQCD, as embodied in the DGLAP
equations, is valid at small-$x$, since this is the formalism used for 
determining PDFs. In the present contribution we address the
question of how PDF uncertainties at low $x$ affect the SM processes 
of $W$ and $Z$ production at the LHC.

The major source of information on low-$x$ physics in the last decade has been
the HERA data. One of the most striking results of HERA was observation
of an unexpected rise
of the $F_2$ structure function at low-$x$.  
The interpretation of the rise in $F_2$, 
in the DGLAP formalism, attributes it to a strong rise in the 
gluon distribution function at low-$x$, since the gluon drives the sea 
distributions by $g \to q \bar{q}$ splitting. 
In fact, the DGLAP equations predict that, at high 
$Q^2 (\geqsim 100$GeV$^2$), 
a steep rise of the gluon and the sea at low-$x$ will evolve from flat input
shapes at a low $Q^2(\sim 4$GeV$^2$). Nevertheless the rise was unexpected, 
firstly, because most theoreticians expected any such tendency
 to be tamed either by screening effects, or by gluon recombination 
at high gluon density. Secondly, because the rise was 
already present for low $Q^2(\sim 1-2$GeV$^2$) - even lower than the 
conventional starting scale for QCD evolution. Hence
the observation of the rise led to excitement in a somewhat orthogonal section
 of the theoretical community, where a steep rise at low $Q^2$ had been 
predicted in the BFKL formalism, which resums diagrams involving
$ln(1/x)$. Such resummations are not part of the 
conventional DGLAP $ln(Q^2)$ summations.  

\begin{figure}[tbp] 
\centerline{
\epsfig{figure=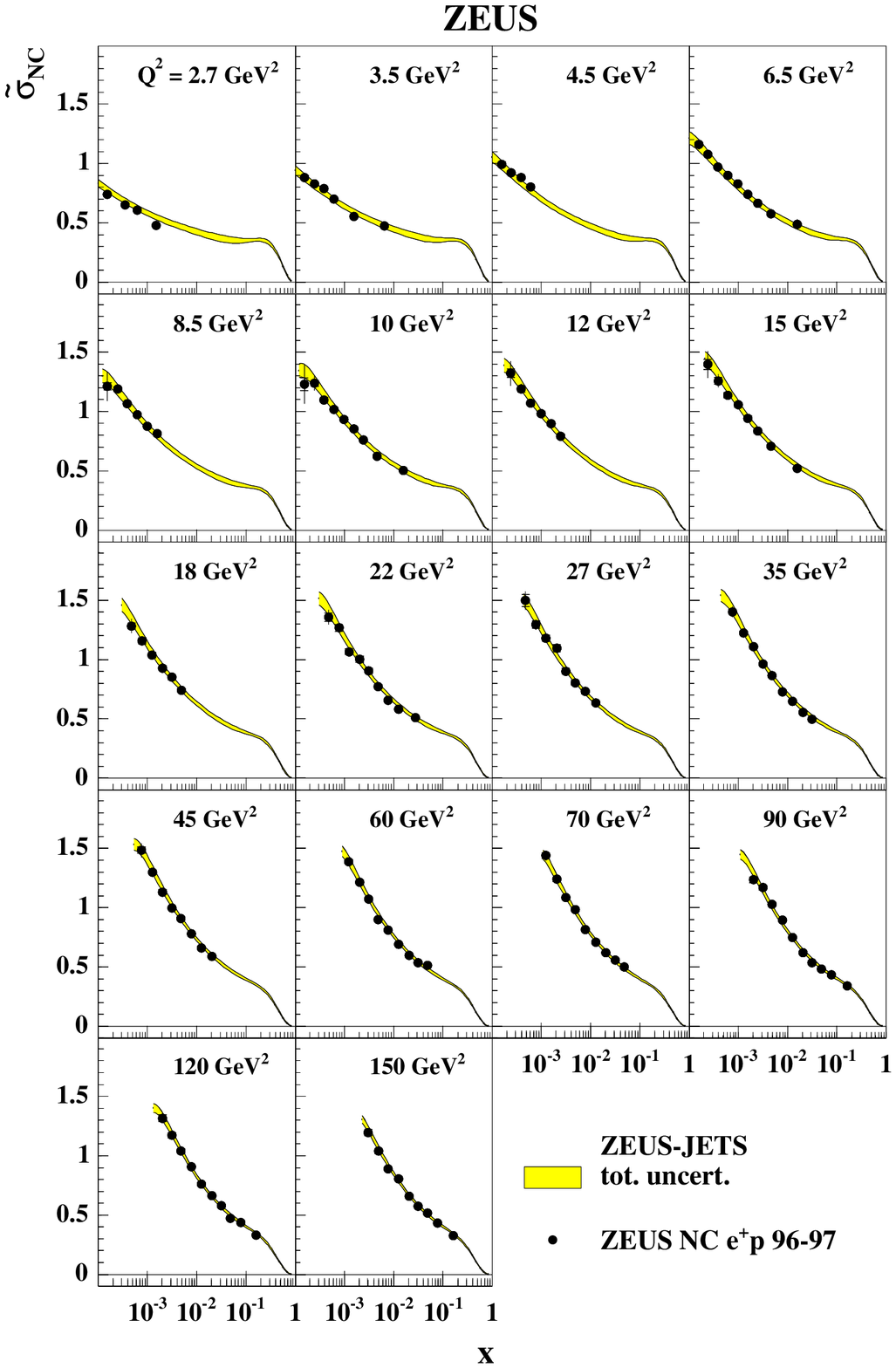,width=0.45\textwidth}
\epsfig{figure=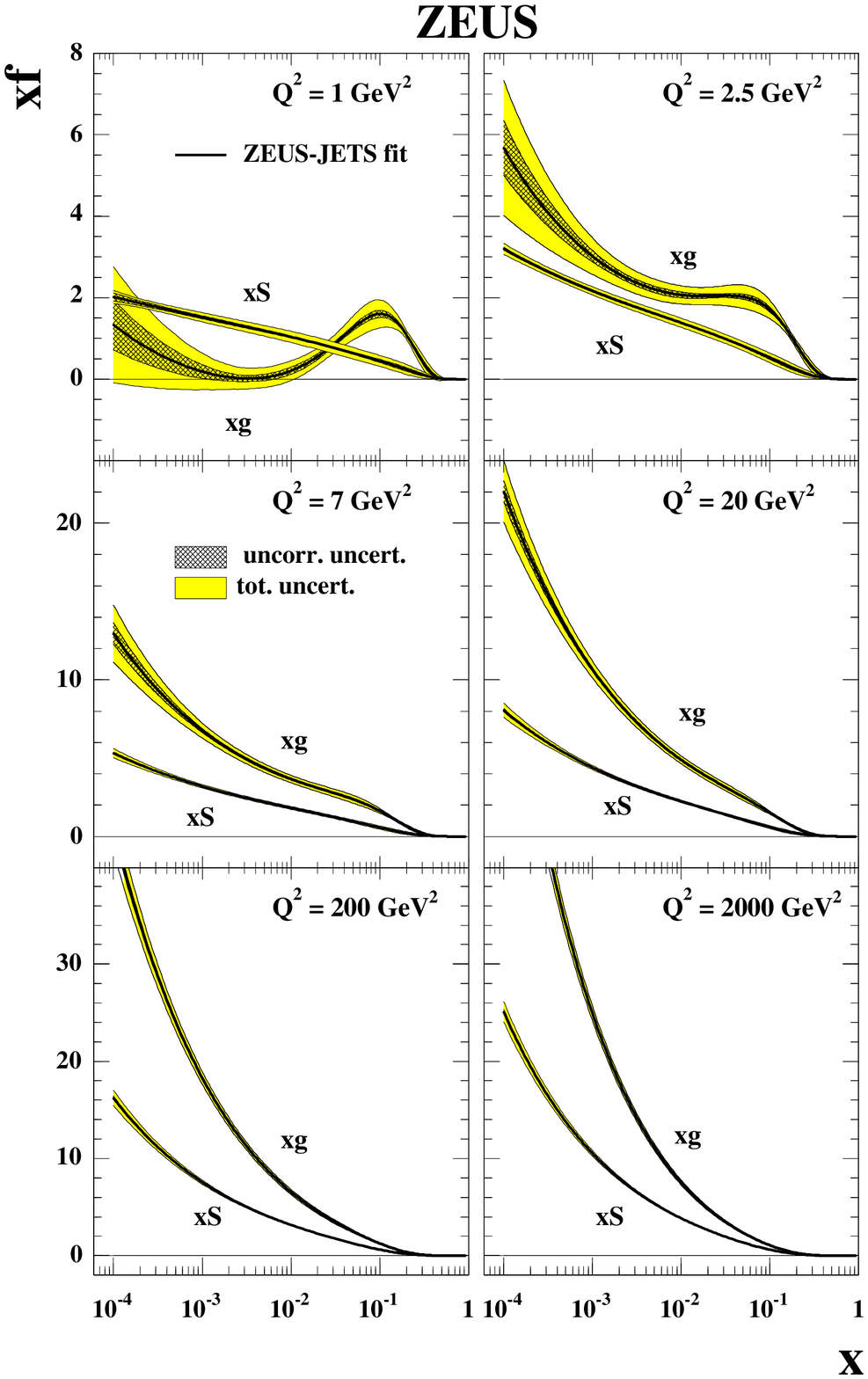,width=0.4\textwidth}}
\caption {Left side: ZEUS data on $F_2$ showing the rise at low $x$, for 
various $Q^2$. Right side: the gluon and sea PDFs extracted from these data in
the ZEUSJETS PDF fit, for various $Q^2$, illustrating the turnover of the gluon
PDF at low $Q^2$ }
\label{fig:NCglusea}
\end{figure}

However, even though the observation of a rise of $F_2$ at low $x$ and 
low $Q^2$ defied conventional prejudice, it can be accommodated within the
conventional DGLAP formalism provided sufficiently flexible input shapes are 
used at a low enough input scale (now taken to be $Q^2 \sim 1$GeV$^2$). In fact
it turns out that whereas the input sea distribution is still rising at 
low-$x$, the input gluon distribution has turned over to become valence-like, 
and is even allowed to become negative in some parametrisations. 
Fig.~\ref{fig:NCglusea} illustrates this behaviour in the sea and gluon PDFs,
together with the data used to extract them.
This counter intuitive behaviour has 
led many QCD theorists to believe that the conventional formalism is in need 
of extension~\cite{dcs}. The contribution of R. Ball to these 
proceedings describes modern work in this area. The present 
contribution is concerned with how well the PDFs are known at low-$x$, within 
the conventional framework, and how this affects the predictions for $W$ and 
$Z$ production at the LHC. These processes have been suggested as 
`standard-candles' for the measurement of luminosity because their 
cross-sections are `well known'. In the present contribution we investigate to 
what extent this is really true- and what might be done about it.
  
\section{$W$ AND $Z$ PRODUCTION AT THE LHC} 
\label{sec:lowx;amcs_wzpred}

At leading order (LO), $W$ and $Z$ production occur by the process, 
$q \bar{q} \rightarrow W/Z$.
Consulting Fig.~\ref{fig:kin/pdfs}, we see that at central rapidity, 
the participating partons have small momentum fractions, $x \sim 0.005$, and
over the meaurable rapidity range, $|y| < 2.4$, 
$x$ values remain in the range, $5.10^{-4} < x < 0.05$. 
Thus, in contrast to the situation at the Tevatron, 
the scattering is happening dominantly between sea quarks and antiquarks. 
Furthermore, the high scale of the process $Q^2 = M^2 \sim 10,000$GeV$^2$ 
ensures that the gluon is the dominant 
parton as also illustrated in Fig.~\ref{fig:kin/pdfs}, where the PDFs for all 
parton flavours 
are shown for $Q^2 = \sim 10,000$GeV$^2$. Hence the sea quarks have 
mostly been generated by the flavour blind $g \to q \bar{q}$ splitting 
process. Thus the precision of our knowledge of $W$ and $Z$ cross-sections at 
the LHC is crucially dependent on the uncertainty on 
the momentum distribution of the gluon at low-$x$. This is where the HERA data
come in. 

\begin{figure}[tbp] 
\centerline{
\epsfig{figure=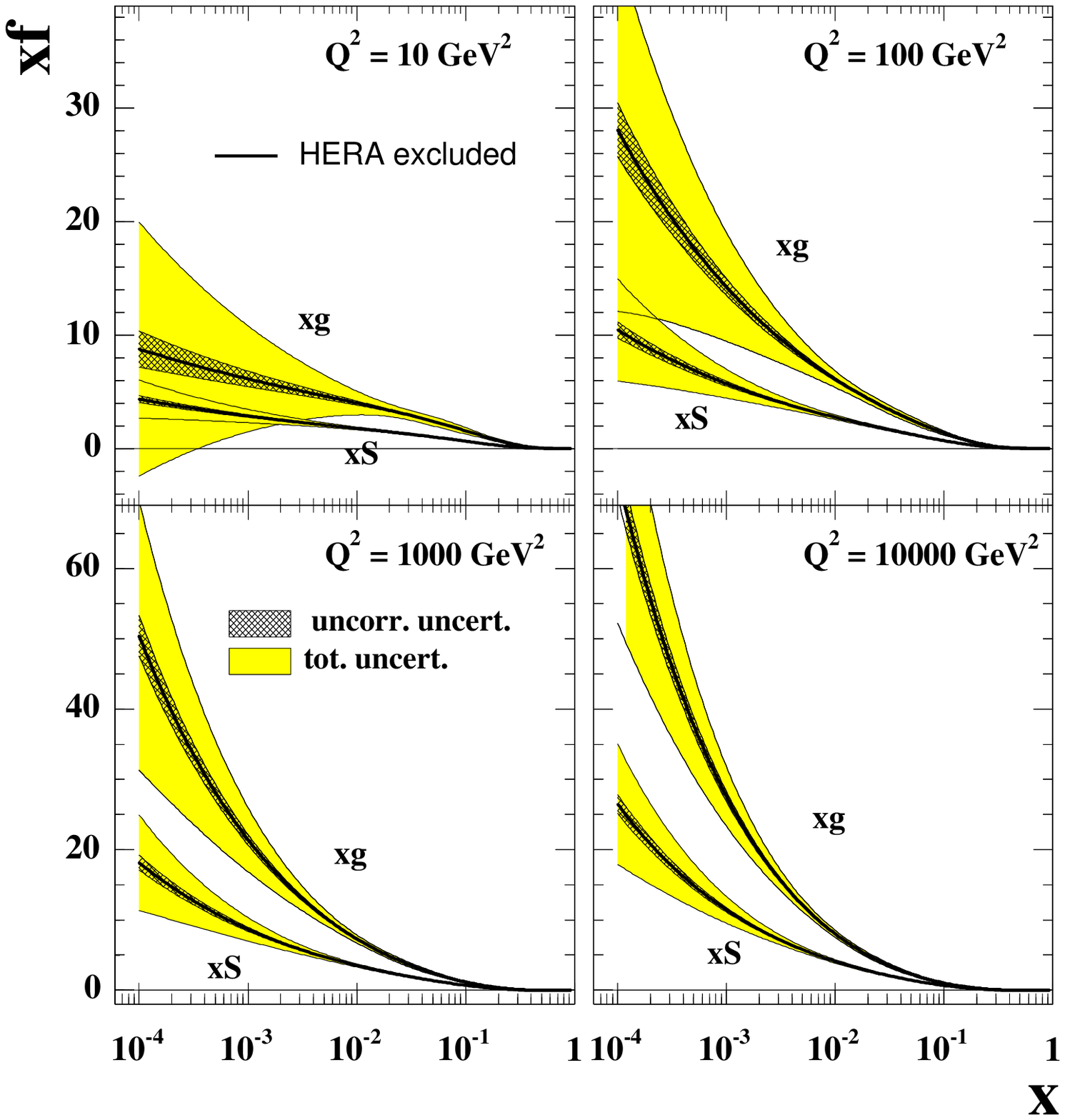,width=0.5\textwidth}
\epsfig{figure=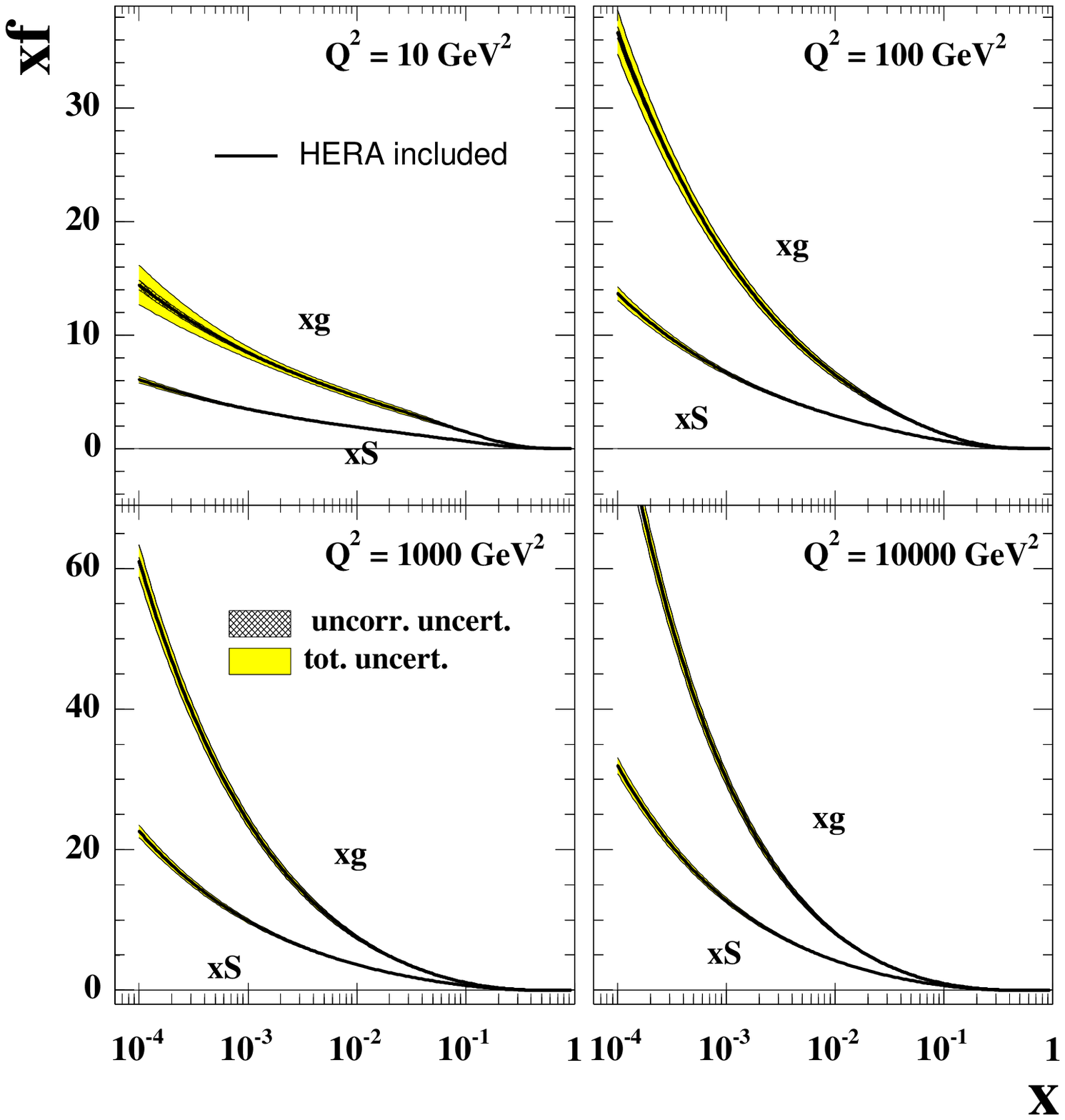,width=0.5\textwidth}}
\caption {Sea ($xS$) and gluon ($xg$) PDFs, as a function of $x$, for various 
$Q^2$ values: left plot; from the ZEUS-S global PDF analysis not including
 HERA data; right plot: from the ZEUS-S global PDF analysis including HERA data}
\label{fig:pre/postPDFs}
\end{figure}

Fig.~\ref{fig:pre/postPDFs} shows the sea and gluon PDFs (and their 
uncertainties) extracted from an
NLO QCD PDF fit analysis to world data on deep inelastic 
scattering, before and after HERA data are included. 
The latter fit is the ZEUS-S global fit~\cite{zeus-s}, whereas the former is 
a fit using the same fitting analysis but leaving out the ZEUS data. The full 
PDF uncertainties for both fits are calculated from the eigenvector PDF sets 
of the ZEUS-S analysis using LHAPDF~\cite{LHAPDF}. The improvement in the 
level of uncertainty is striking.
 
\begin{figure}[tbp] 
\vspace{-1.0cm}
\centerline{
\epsfig{figure=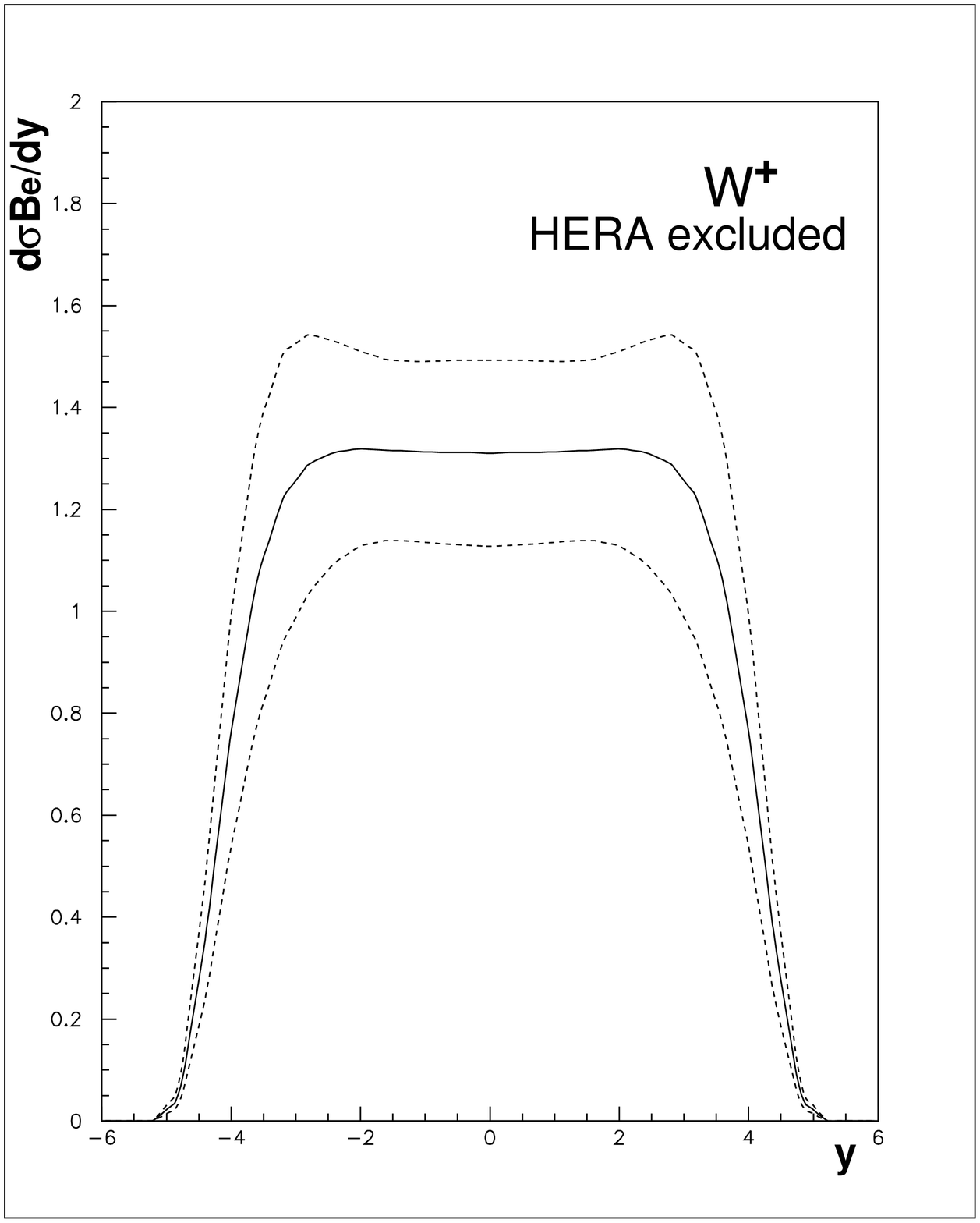,width=0.3\textwidth,height=4.5cm}
\epsfig{figure=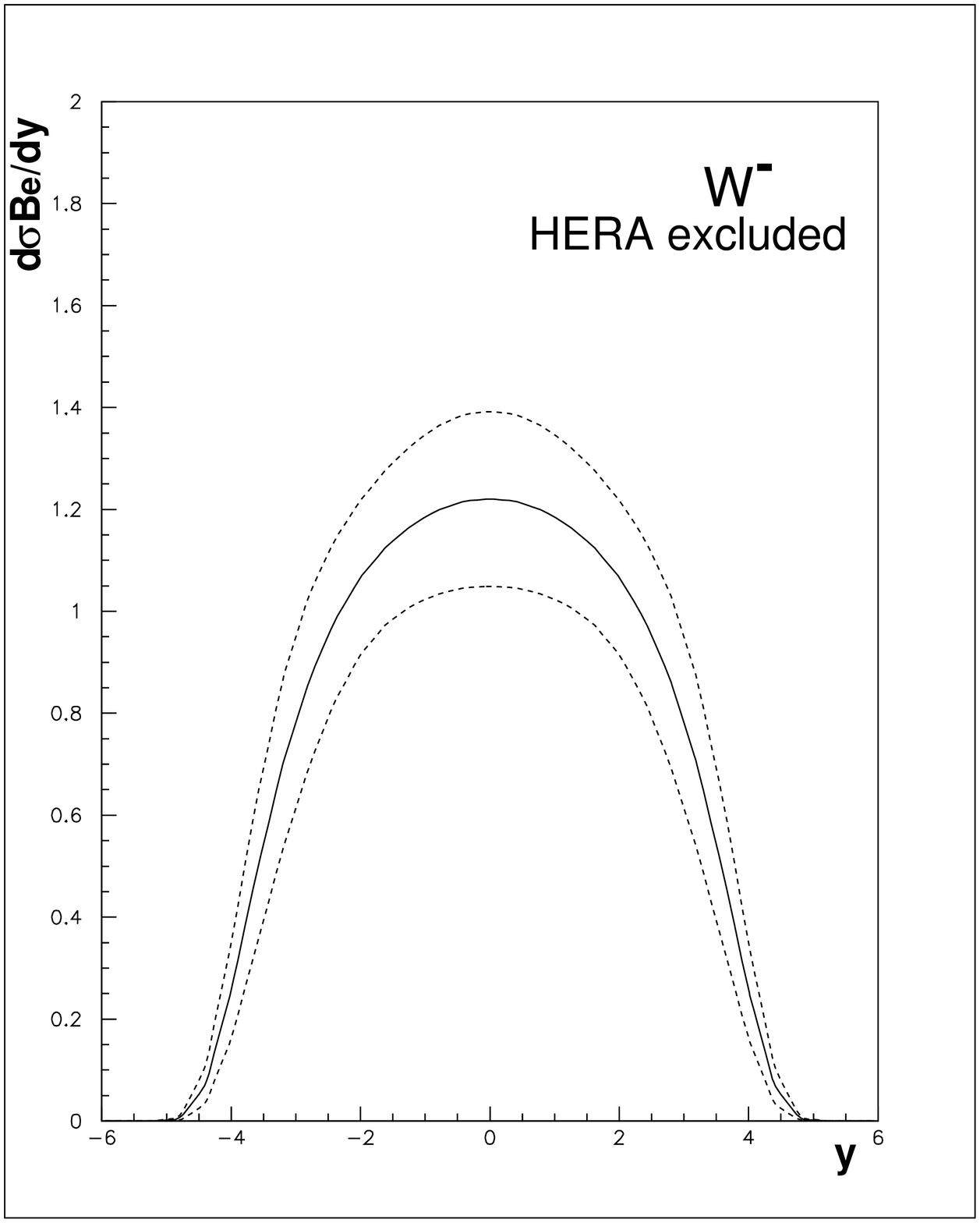,width=0.3\textwidth,height=4.5cm}
\epsfig{figure=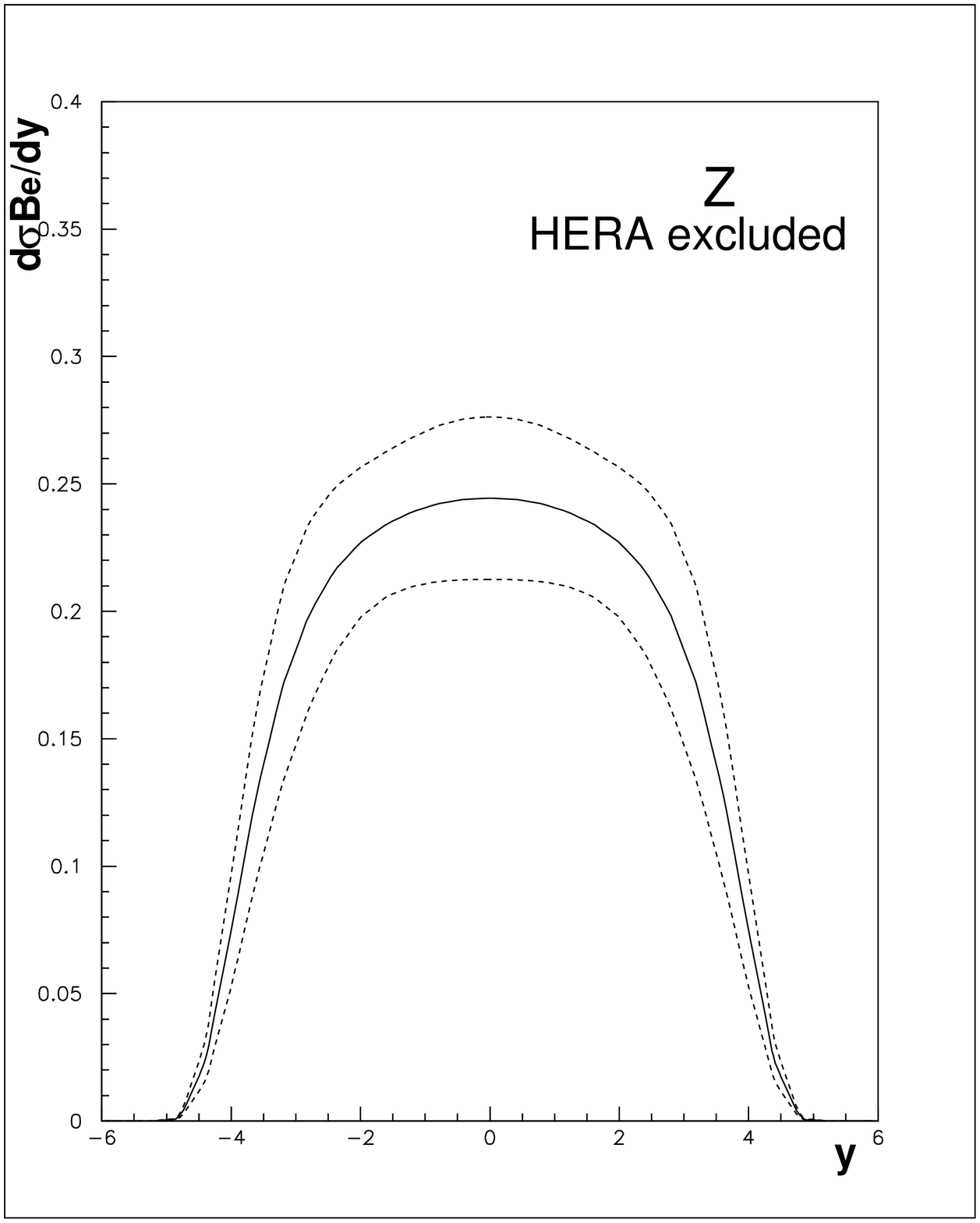,width=0.3\textwidth,height=4.5cm} 
}
\centerline{
\epsfig{figure=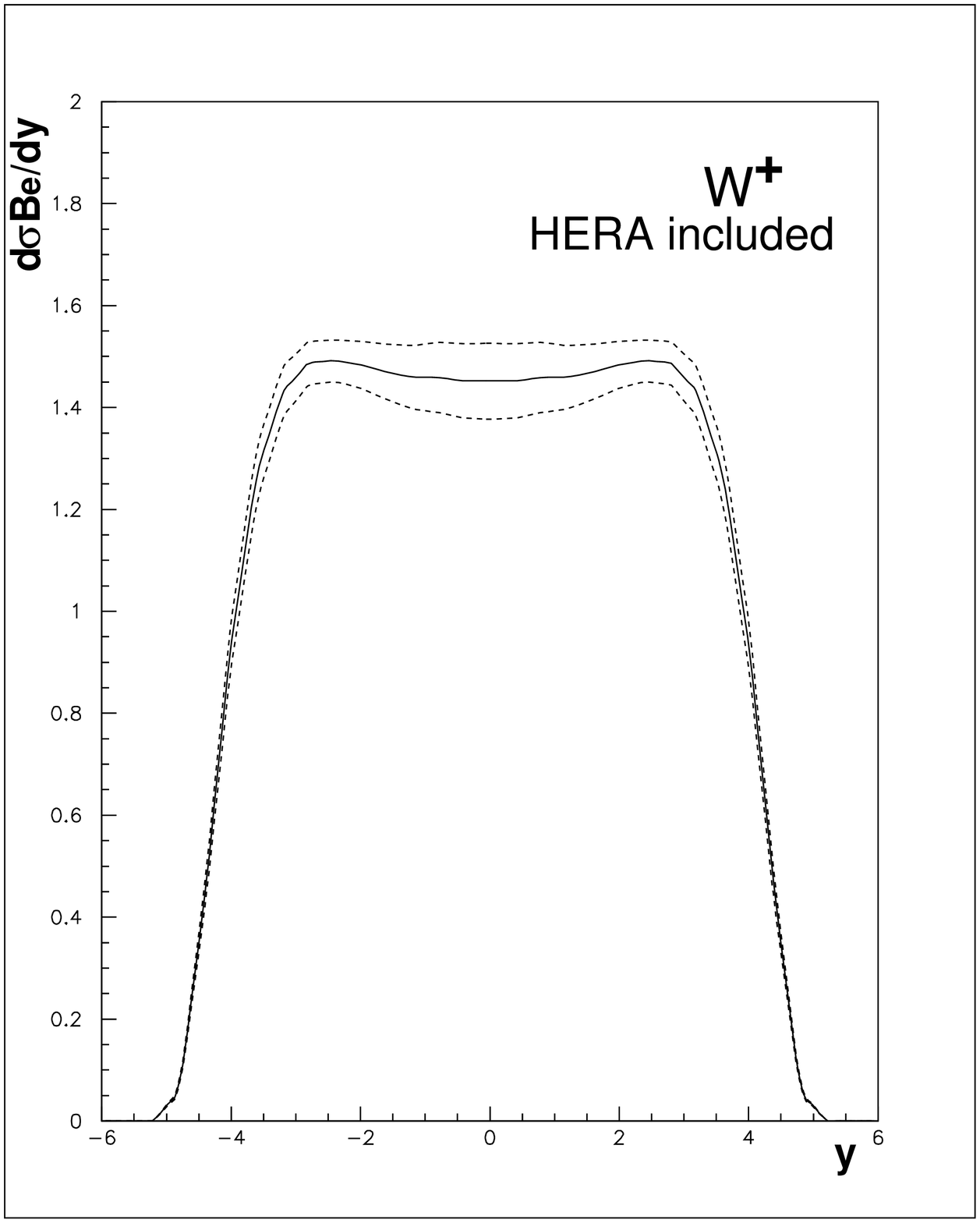,width=0.3\textwidth,height=4.5cm}
\epsfig{figure=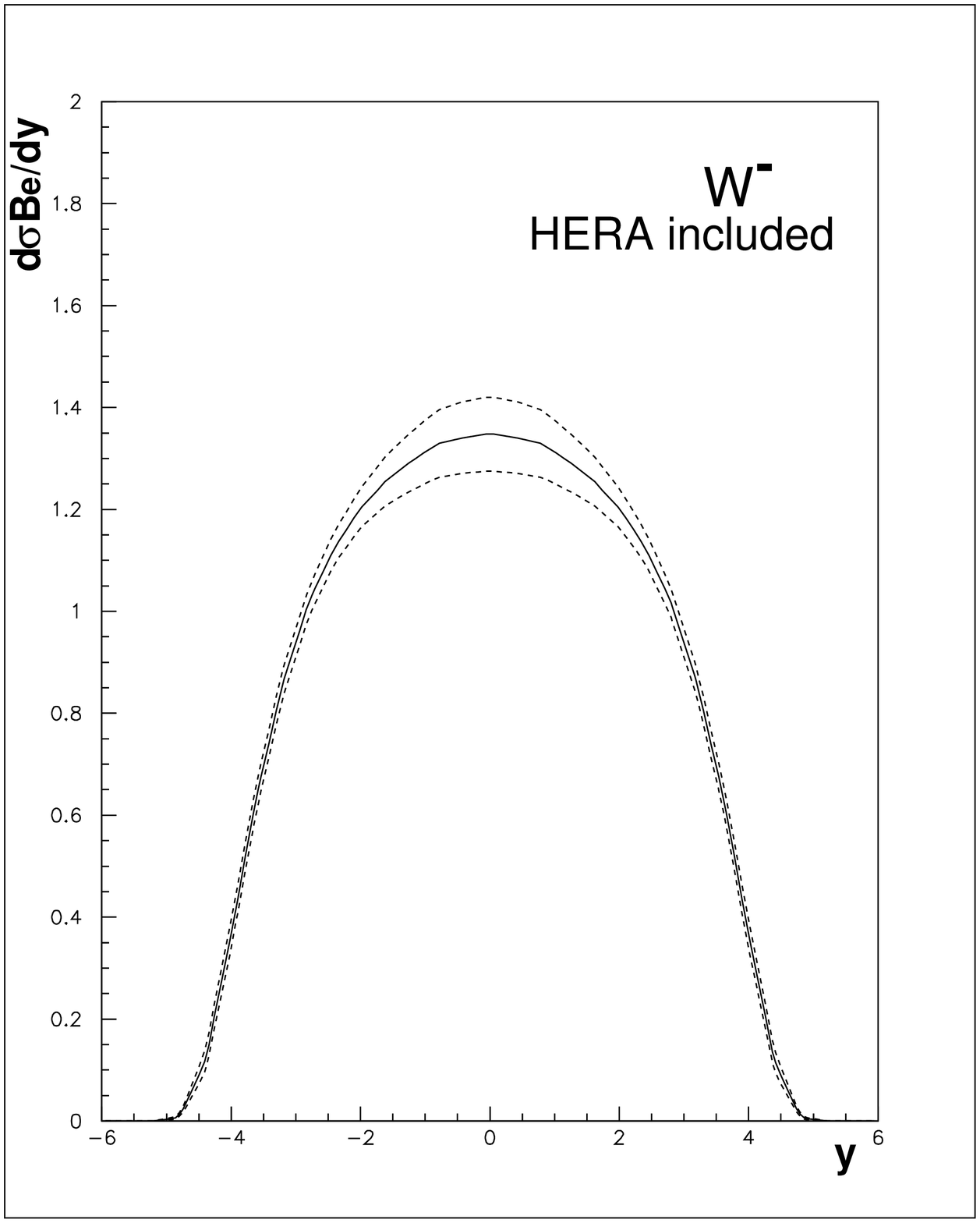,width=0.3\textwidth,height=4.5cm}
\epsfig{figure=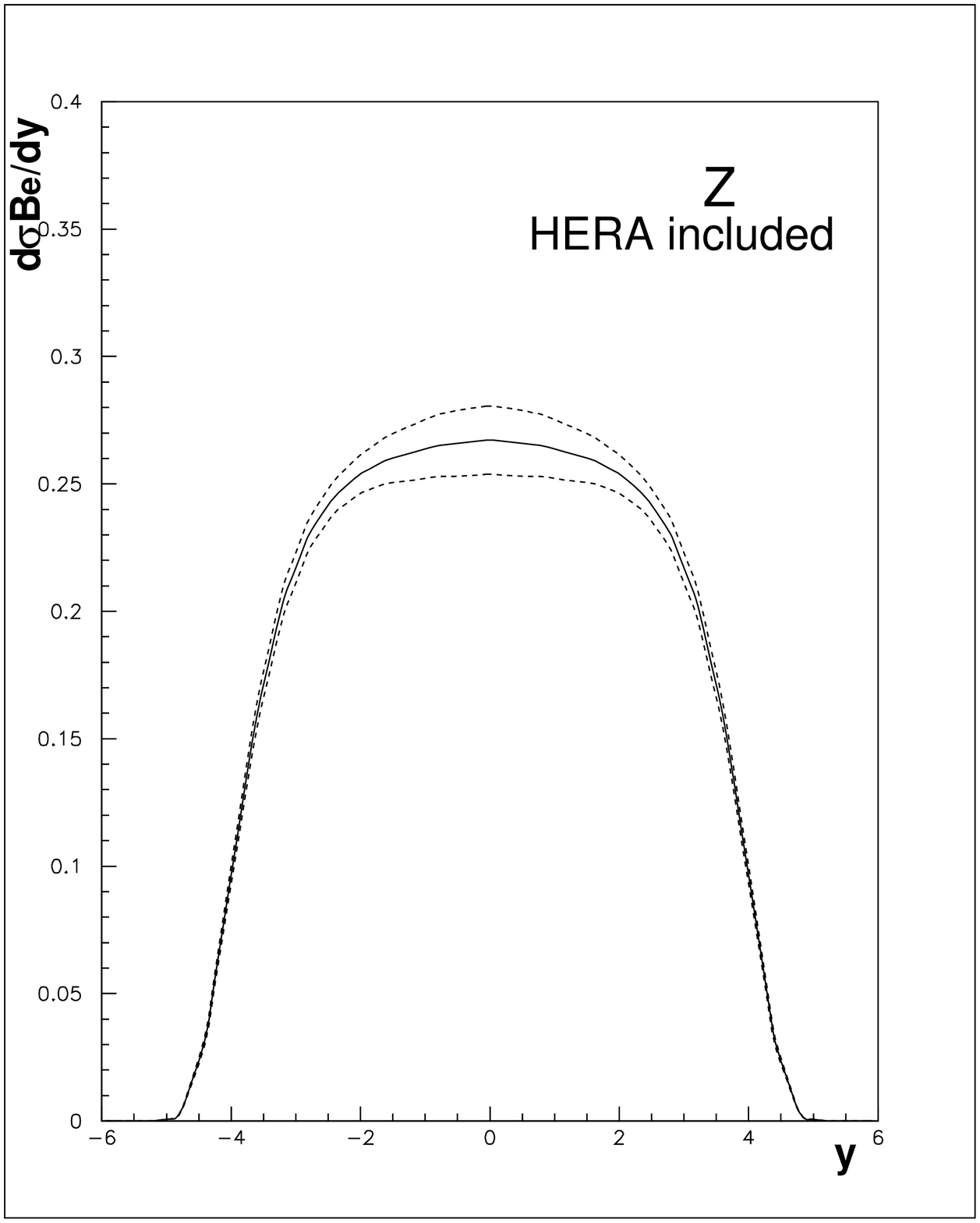,width=0.3\textwidth,height=4.5cm}
}
\caption {LHC $W^+,W^-,Z$ rapidity distributions and their PDF uncertainties: Top Row: from the ZEUS-S 
global PDF analysis
not including HERA data; left plot $W^+$; middle plot $W^-$; right plot $Z$: Bottom Row: from the ZEUS-S
global PDF analysis including HERA data; left plot $W^+$; middle plot $W^-$; right plot $Z$}
\label{fig:WZrapFTZS13}
\end{figure}

Fig.~\ref{fig:WZrapFTZS13} illustrates how this improved knowledge of the gluon
and sea distributions has improved our knowledge of $W$ and $Z$ 
cross-sections.
It shows $W$ and $Z$ rapidity spectra predicted using the PDFs extracted from
the global PDF fit which does not include the HERA data, compared to 
those extracted from the similar global PDF fit which does include HERA data. 
The corresponding predictions for the $W/Z$ cross-sections, 
decaying to the lepton decay mode, are summarised in Table~\ref{tab:datsum}.
\begin{table}[t]
\centerline{\small
\begin{tabular}{llllcccc}\\
 \hline
PDF Set  & $\sigma(W^+).B(W^+ \rightarrow l^+\nu_l)$ & $\sigma(W^-).B(W^- \rightarrow l^-\bar{\nu}_l)$ & 
$\sigma(Z).B(Z \rightarrow l^+ l^-)$\\
 \hline
 ZEUS-S no HERA  & $10.63 \pm 1.73 $~nb & $7.80 \pm 1.18 $~nb & $1.69 \pm 0.23$~nb \\
 ZEUS-S  & $12.07 \pm 0.41 $~nb & $8.76 \pm 0.30 $~nb & $1.89 \pm 0.06$~nb\\
 CTEQ6.1 & $11.66 \pm 0.56 $~nb & $8.58 \pm 0.43 $~nb & $1.92 \pm 0.08$~nb\\
 MRST01 & $11.72 \pm 0.23 $~nb & $8.72 \pm 0.16 $~nb & $1.96 \pm 0.03$~nb\\
 \hline\\
\end{tabular}}
\caption{LHC $W/Z$ cross-sections for decay via the lepton mode, for various PDFs}
\label{tab:datsum}
\end{table}
The uncertainties in the predictions for these cross-sections have decreased 
from $\sim 16\%$ pre-HERA to $\sim 3.5\%$ post-HERA. There could clearly have 
been no talk of using these processes as standard candle processes, 
without the HERA data.

\begin{figure}[tbp] 
\centerline{
\epsfig{figure=my_wprap_lha_zs13.eps,width=0.3\textwidth,height=4.5cm}
\epsfig{figure=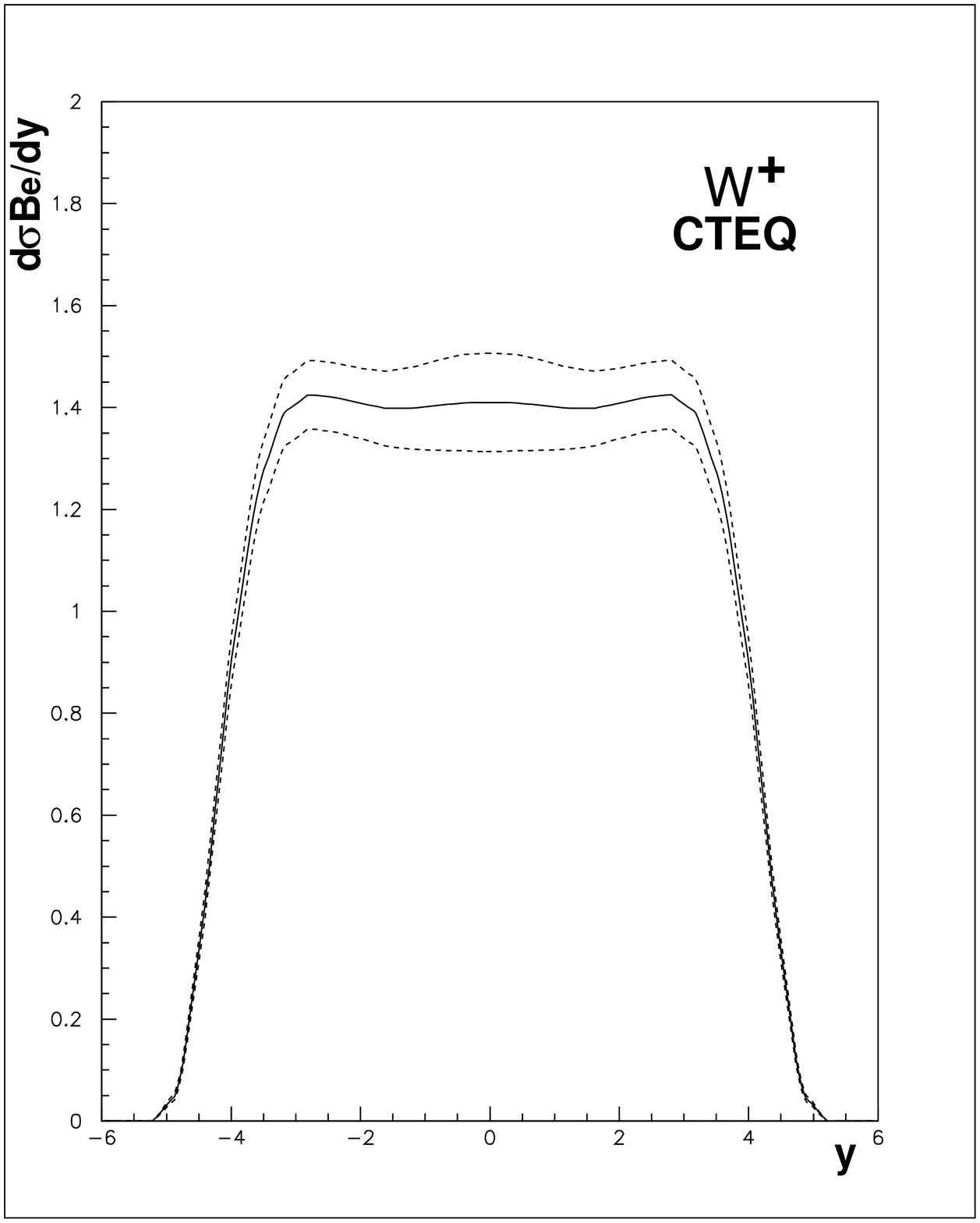,width=0.3\textwidth,height=4.5cm}
\epsfig{figure=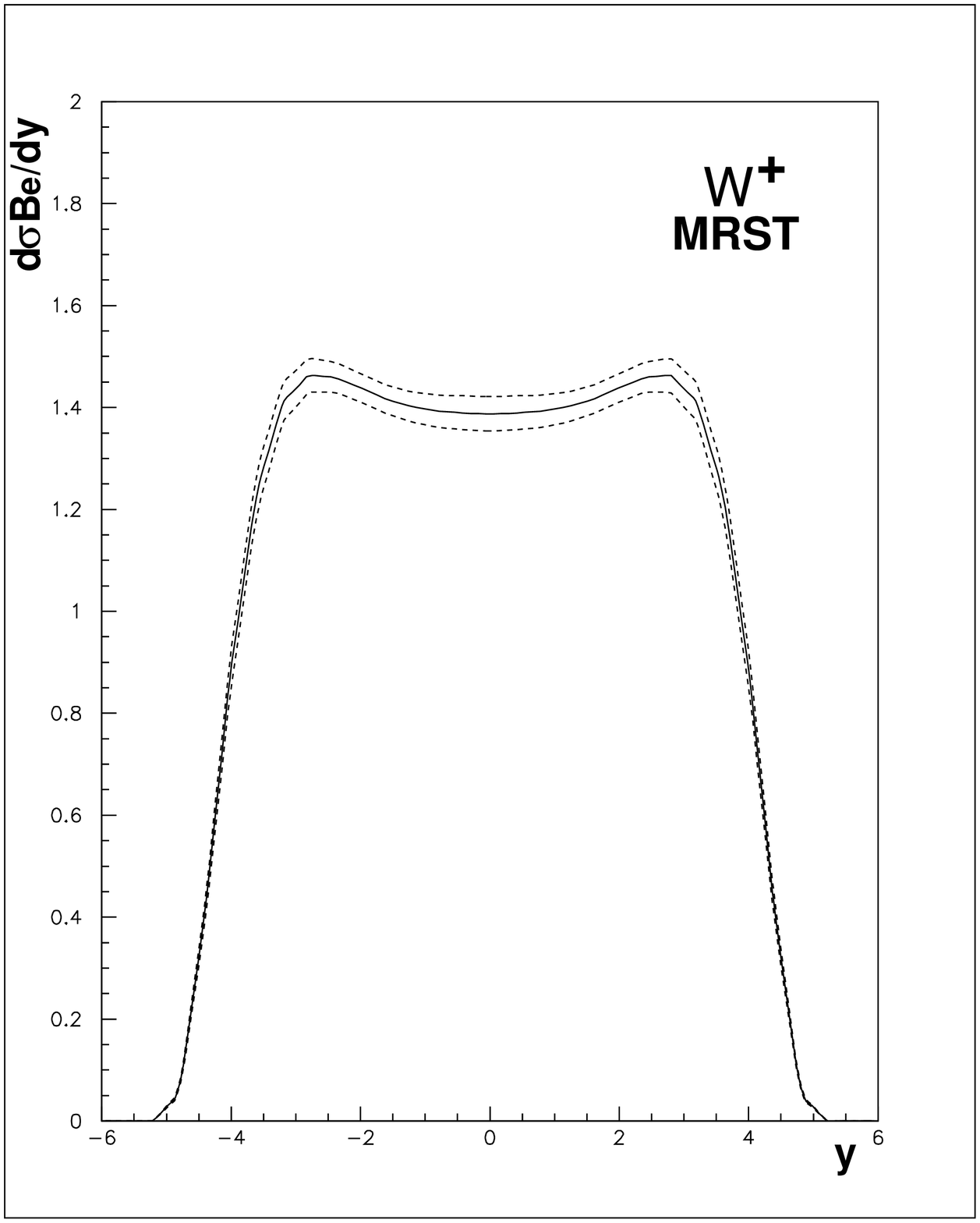,width=0.3\textwidth,height=4.5cm}
}
\caption {LHC $W^+$ rapidity distributions and their PDF uncertainties:
 left plot, ZEUS-S PDFs; middle plot, CTEQ6.1 PDFs;
right plot: MRST01 PDFs.}
\label{fig:mrstcteq}
\end{figure}

The post-HERA level of precision illustrated in Fig.~\ref{fig:WZrapFTZS13} 
is taken for granted in modern analyses. However, when 
considering the PDF uncertainties on the Standard Model (SM) predictions it 
is necessary not 
only to consider the uncertainties of a particular PDF analysis, but also to 
compare PDF analyses. Fig.~\ref{fig:mrstcteq} compares the predictions for 
$W^+$ production for the ZEUS-S PDFs with those of 
the CTEQ6.1\cite{cteq} PDFs and the MRST01\cite{mrst} PDFs\footnote{MRST01 PDFs are used because the 
full error analysis is available for this PDF set.}. 
The corresponding $W^+$ cross-sections, for decay to leptonic mode are 
given in Table~\ref{tab:datsum}.
Comparing the uncertainty at central rapidity, rather 
than the total cross-section, we see that the uncertainty estimates are 
somewhat larger: $\sim 6\%$ for ZEUS-S; $\sim 8\%$ 
for CTEQ6.1M and $\sim 3\%$ for MRST01. 
The difference in the central value between 
ZEUS-S and CTEQ6.1 is $\sim 4\%$. Thus the spread in the predictions of the 
different PDF sets is 
comparable to the uncertainty estimated by the individual analyses. 
Since the measurable rapidity range is restricted to central rapidity it is 
more prudent to use these uncertainty estimates when considering if $W,Z$
cross-sections can be used as luminosity monitors. Comparing the results from
the three PDF extractions it seems reasonable to use generous estimate of the
CTEQ6.1 analysis, $8\%$, as an estimate of how well the luminosity could be
measured, at the present level of uncertainty. We subject this estimate to 
some further reality checks below and in Sec.~\ref{sec:lowx;amcs_reality} and 
we discuss the possibility of improving this estimate with early LHC data in
Sec.~\ref{sec:lowx;amcs_improve}

\begin{figure}[tbp] 
\centerline{
\epsfig{figure=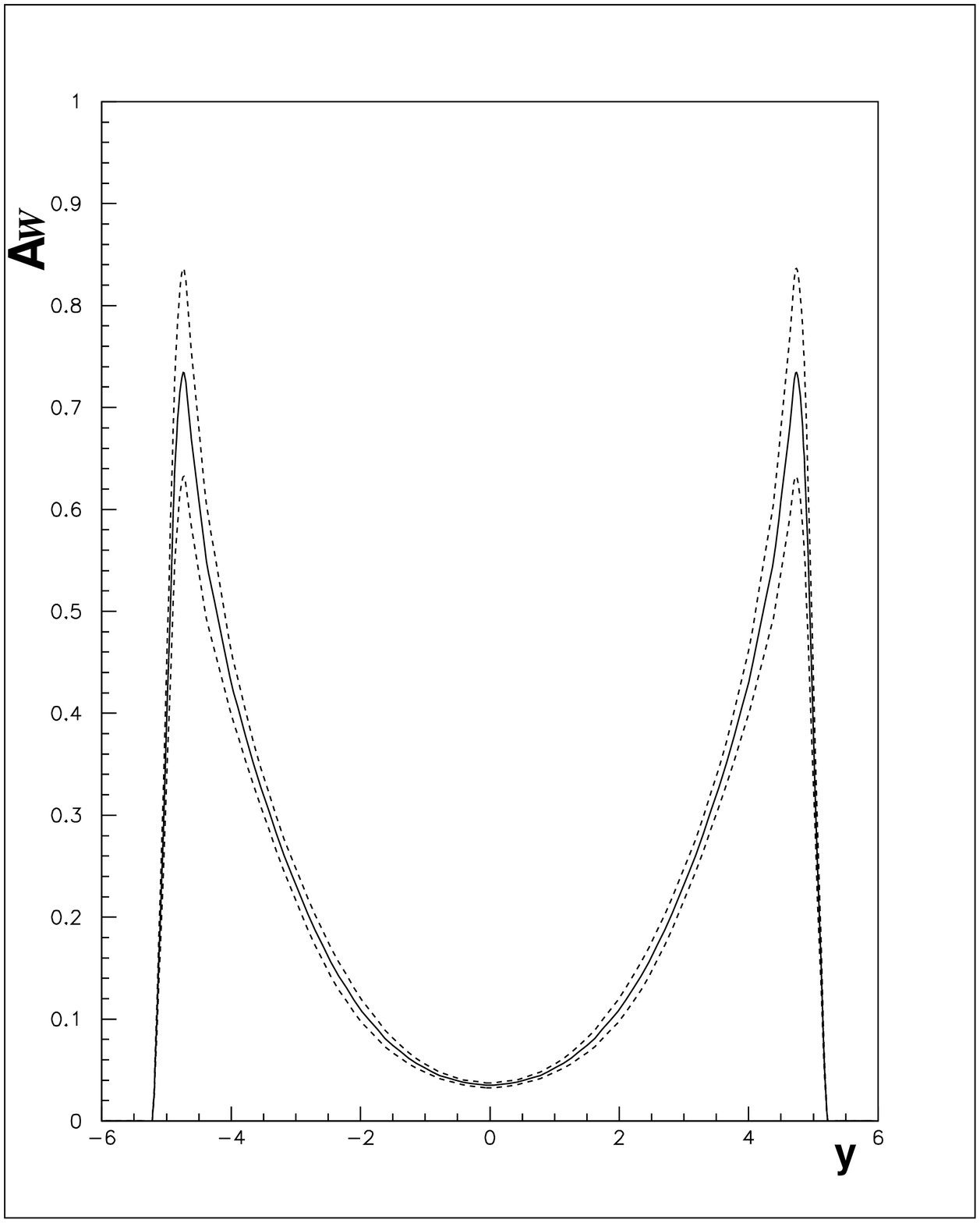,width=0.3\textwidth,height=4.5cm}
\epsfig{figure=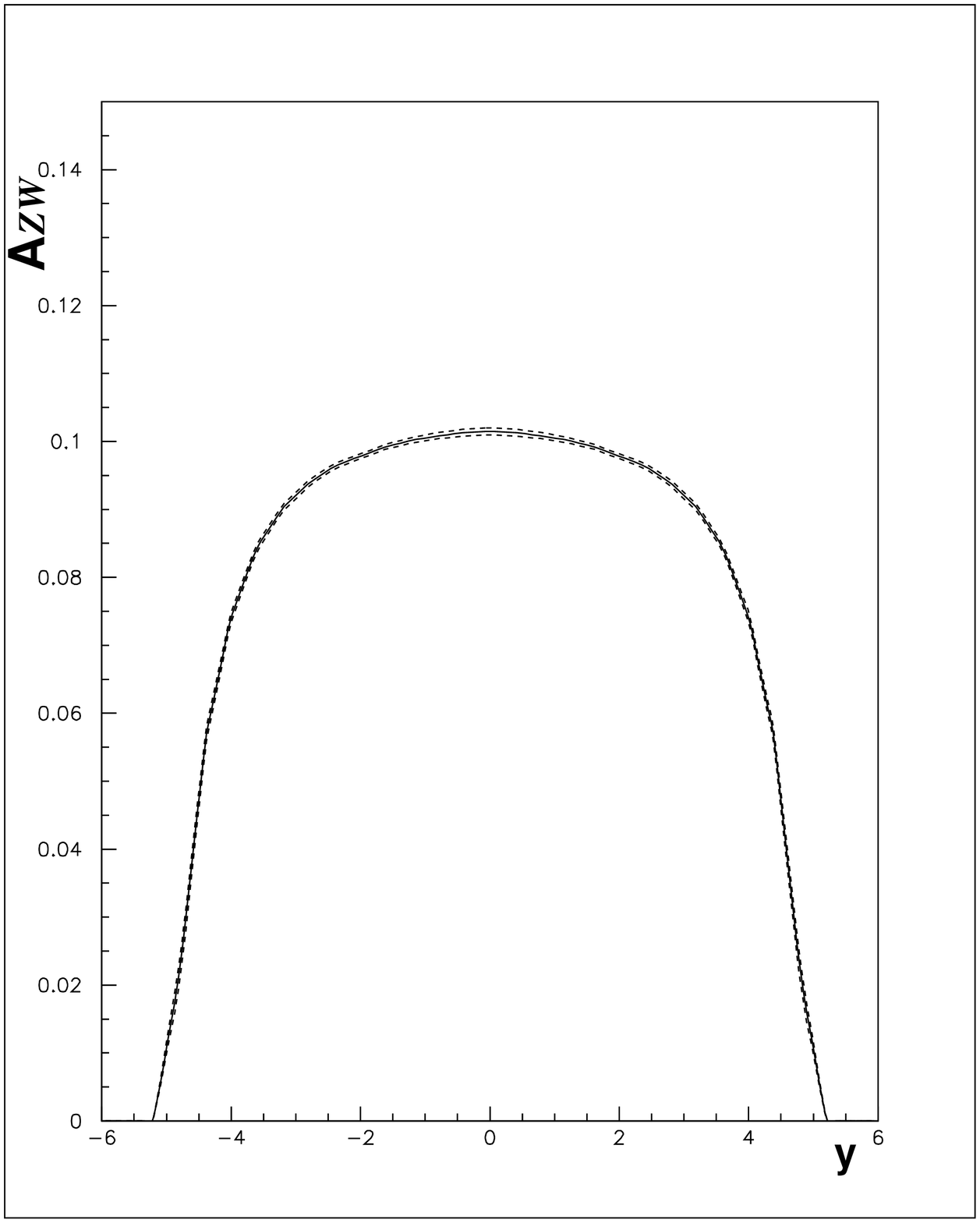,width=0.3\textwidth,height=4.5cm}
\epsfig{figure=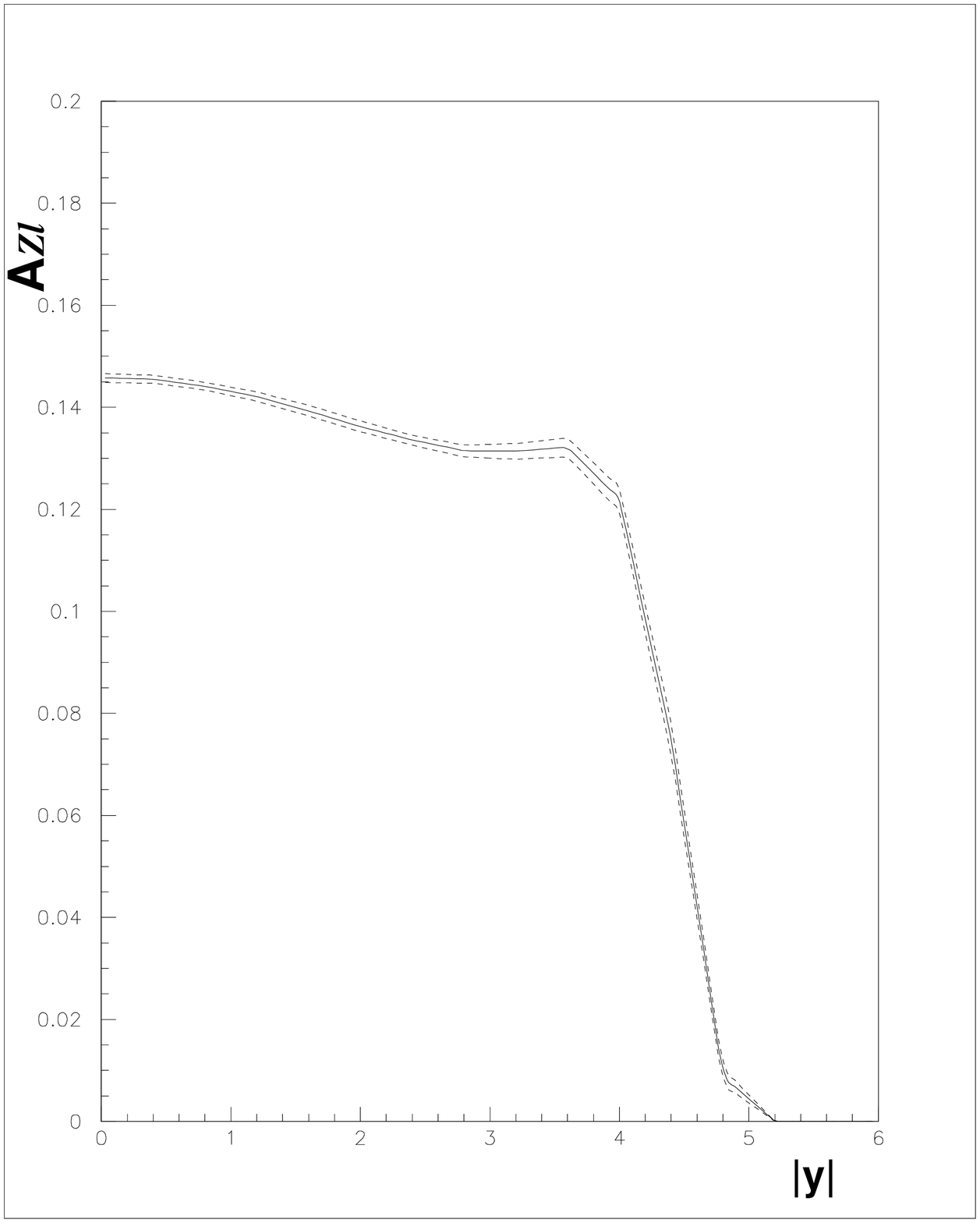,width=0.3\textwidth,height=4.5cm}
}
\centerline{
\epsfig{figure=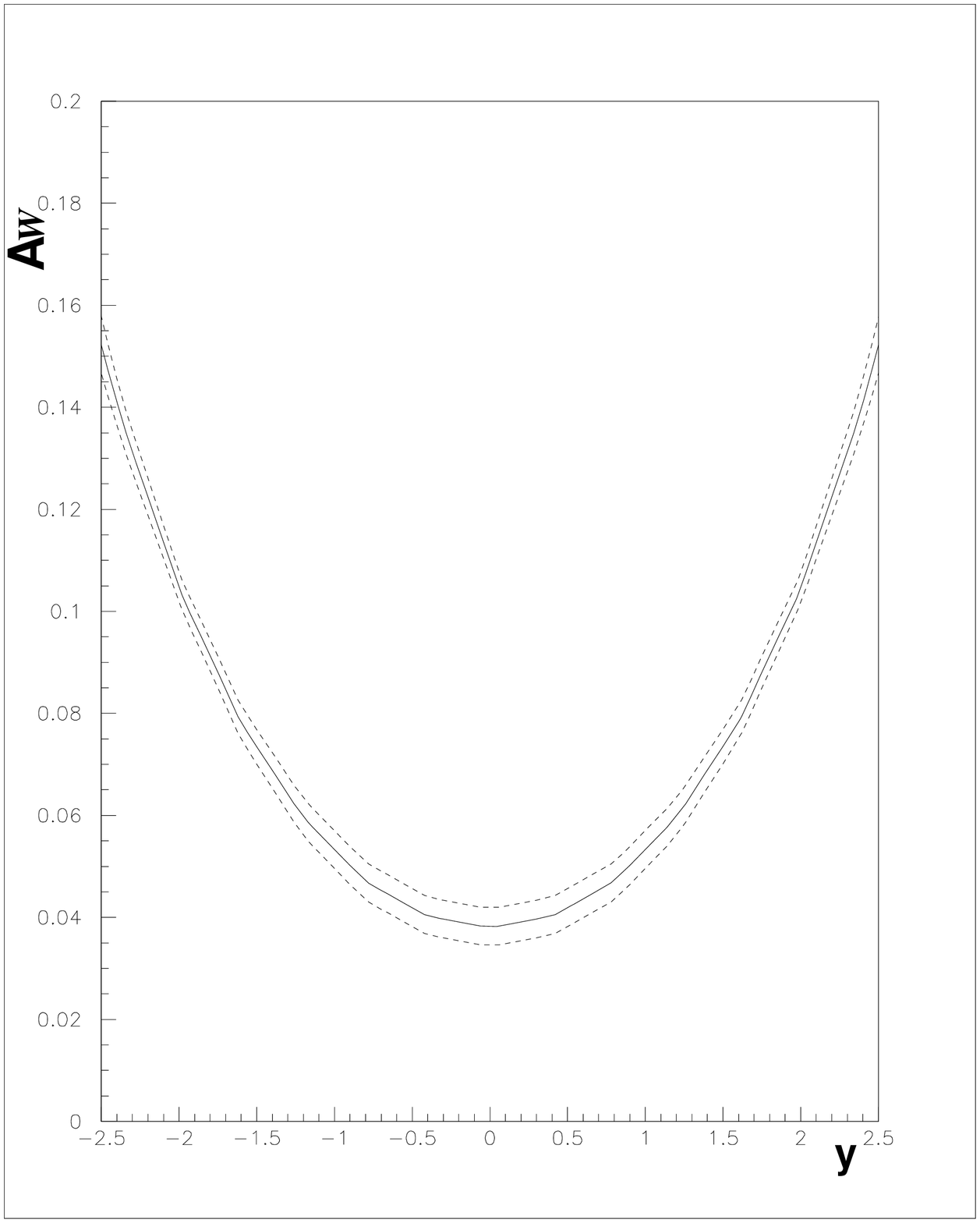,width=0.3\textwidth,height=4.5cm}
\epsfig{figure=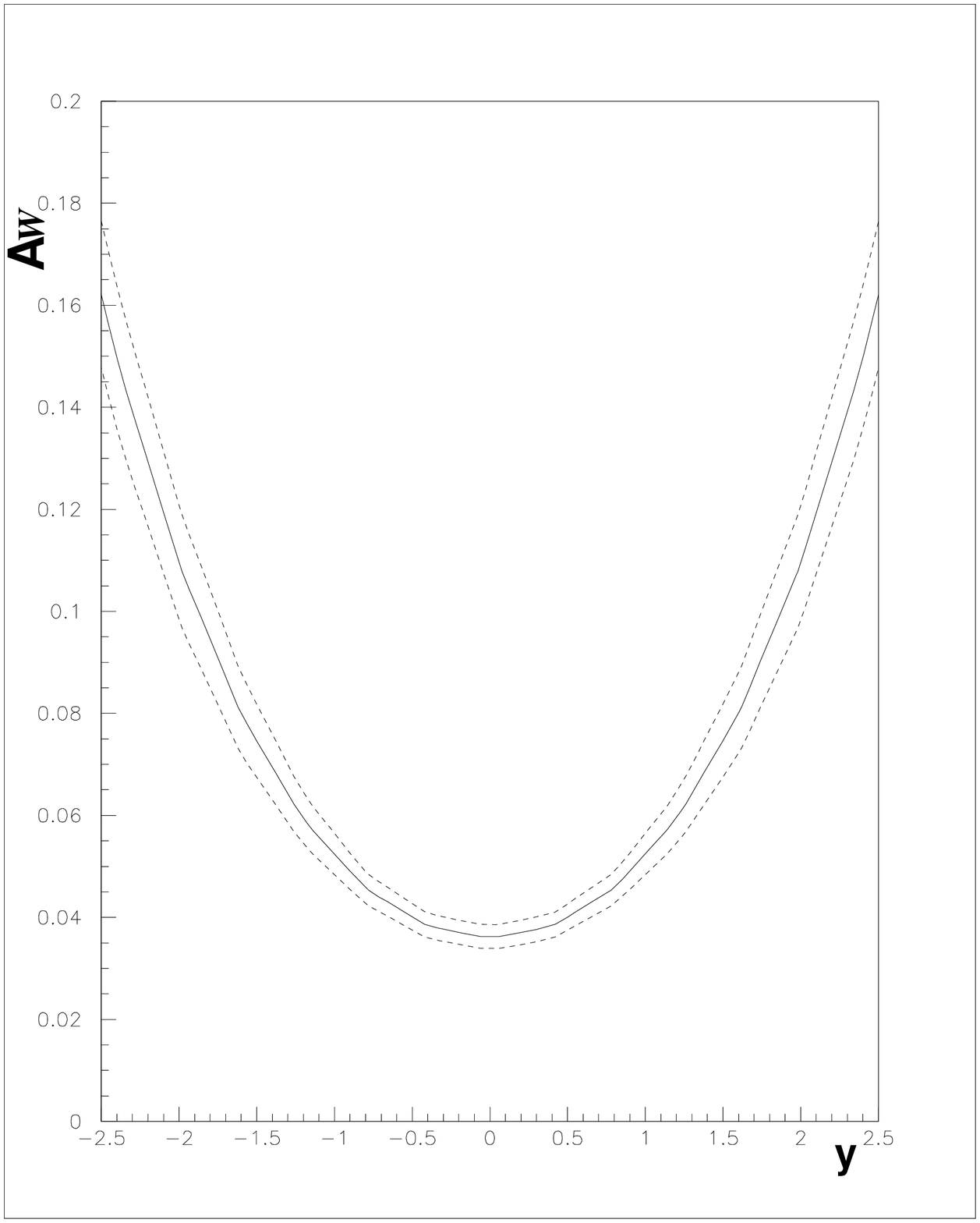,width=0.3\textwidth,height=4.5cm}
\epsfig{figure=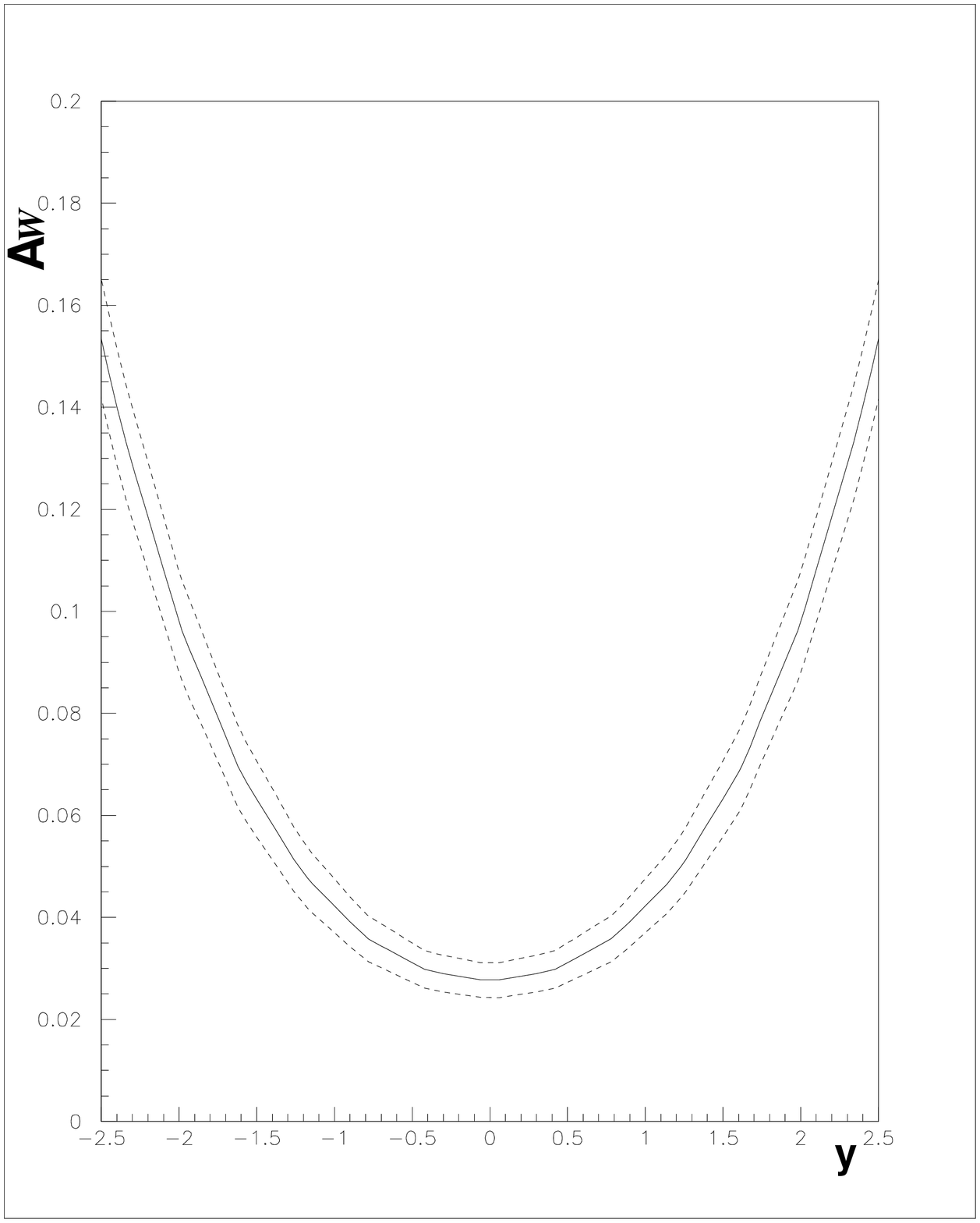,width=0.3\textwidth,height=4.5cm} 
}
\caption {
Top row: predictions from the CTEQ6.1 PDFs: left plot, the $W$ asymmetry, 
$A_W$; middle plot, the ratio, $A_{ZW}$; right plot, the ratio, $A_{Zl}$.  
Bottom row: the $W$ asymmetry, $A_W$, 
within the measurable rapidity range, as predicted using different PDF 
analyses; left plot ZEUS-S; middle plot CTEQ6.1; right plot MRST01.}
\label{fig:awzw}
\end{figure}

Since the PDF uncertainty feeding into the $W^+, W^-$ and $Z$ production is 
mostly coming from the
gluon PDF, for all three processes, there is a correlation in their 
uncertainties, which can be 
removed by taking ratios. The upper half of 
Fig.~\ref{fig:awzw} shows the $W$ asymmetry 
\[A_W = (W^+ - W^-)/(W^+ + W^-).\] for CTEQ6.1 PDFs. 
The PDF uncertainties on the asymmetry at central rapidity are about
$~5\%$, smaller than those on the $W$ spectra themselves, and
a PDF eigenvector decomposition indicates that 
sensitivity to $u$ and $d$ quark flavour distributions is now evident.
Even this residual flavour 
sensitivity can be removed by taking the ratio \[A_{ZW} = Z/(W^+ +W^-)\] 
as also shown in Fig.~\ref{fig:awzw}. 
This quantity is almost independent of PDF uncertainties, which are now as 
small as $~0.5\%$, within the CTEQ6.1 PDF analysis. 

However, as before, 
it is necessary to compare these quantities between different PDF 
analyses. The variation in the predictions for the ratio $A_{ZW}$ between PDF 
analyses (MRST01, ZEUS-S and Alekhin02 PDFs have been compared to CTEQ6.1) is 
outside the PDF uncertainty estimates of the different analyses, but it is 
still only $\sim 5\%$. Hence this ratio could be a used as an SM
benchmark measurement. The ratio, $A_W$, shows
a much more striking difference between MRST01 PDFs and the others. 
This is illustrated in the lower half of 
Fig.~\ref{fig:awzw} for the ZEUS-S, CTEQ6.1 and MRST01 PDFs, 
in the measurable rapidity range. There is a difference of $\sim 25\%$ in the 
predictions.  The origin of this difference between MRST and other PDFs is in the valence spectra. At leading order, the dominant contribution to $A_W$ is
\begin{equation}
 A_W = \frac{u\bar{d} - d \bar{u}}{u\bar{d} + d \bar{u}}.
\end{equation}
At central rapidity, $x \sim 0.005$, for both partons and consequently $\bar{u} \approx \bar{d}$ \footnote{Even if some fairly wild assumptions as to the 
shape of $\bar{d}- \bar{u}$ are made for low $Q^2$, 
the absolute size of 
$\bar{q}$ evolves with $Q^2$ to become very large at $Q^2 = M_W^2$, whereas 
the difference does not evolve, and becomes relatively small.}. Thus
\begin{equation}
 A_W = \frac{u - d}{u + d} = \frac{u_v - d_v}{u_v + d_v +2 \bar{q}}
\end{equation}
and $A_W$ depends on the difference of the valence quarks. 
The quantity 
$u_v- d_v$, is different for MRST and CTEQ, and this difference is outside 
the PDF uncertainty estimates of either
analysis. However, these uncertainty estimates are themselves 
unreliable for valence spectra 
at $x \sim 0.005$, since there is no data on valence quantities at such small 
$x$. The LHC can provide the first such measurement.    

\begin{figure}[tbp] 
\vspace{-1.0cm}
\centerline{
\epsfig{figure=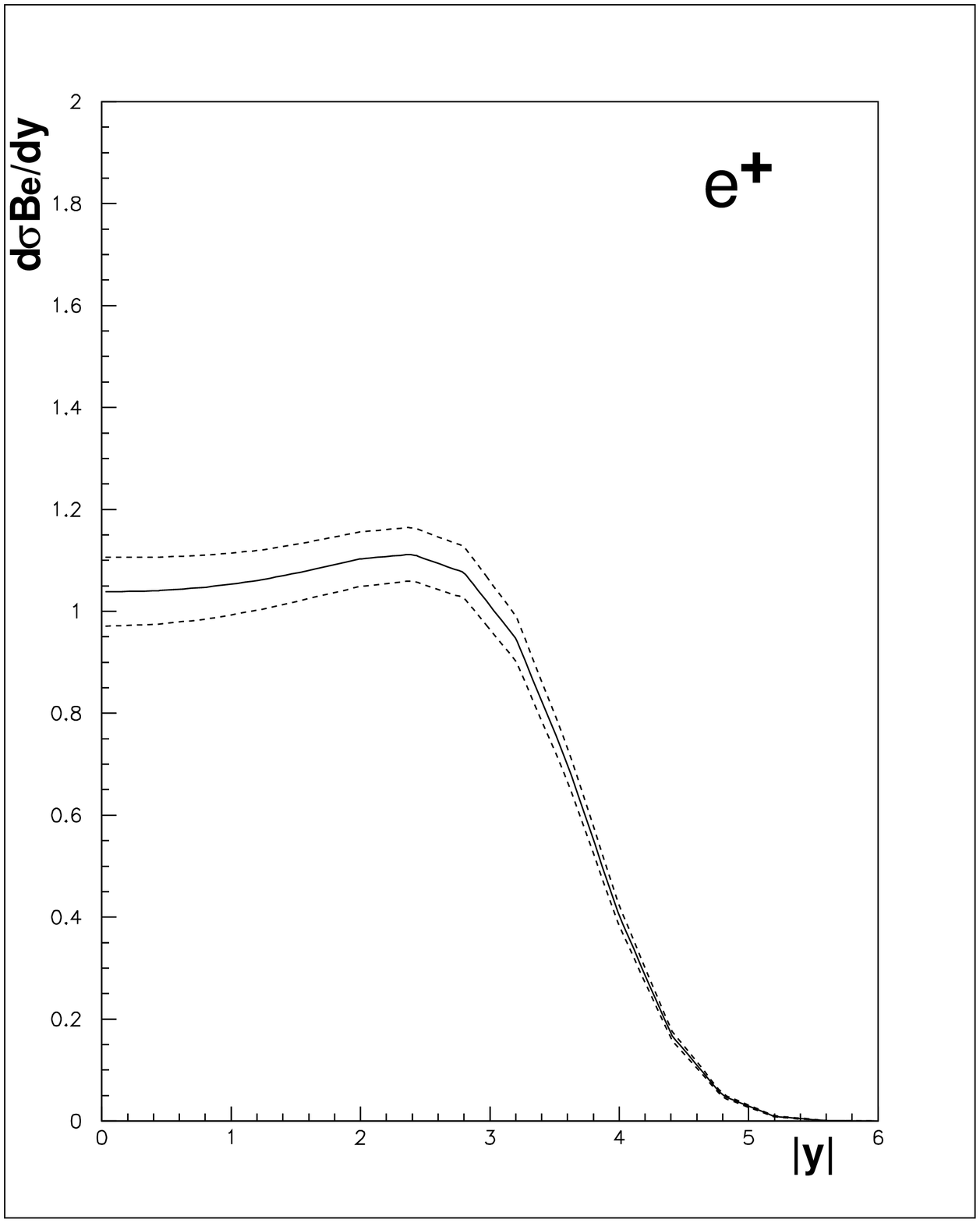,width=0.3\textwidth,height=4.5cm}
\epsfig{figure=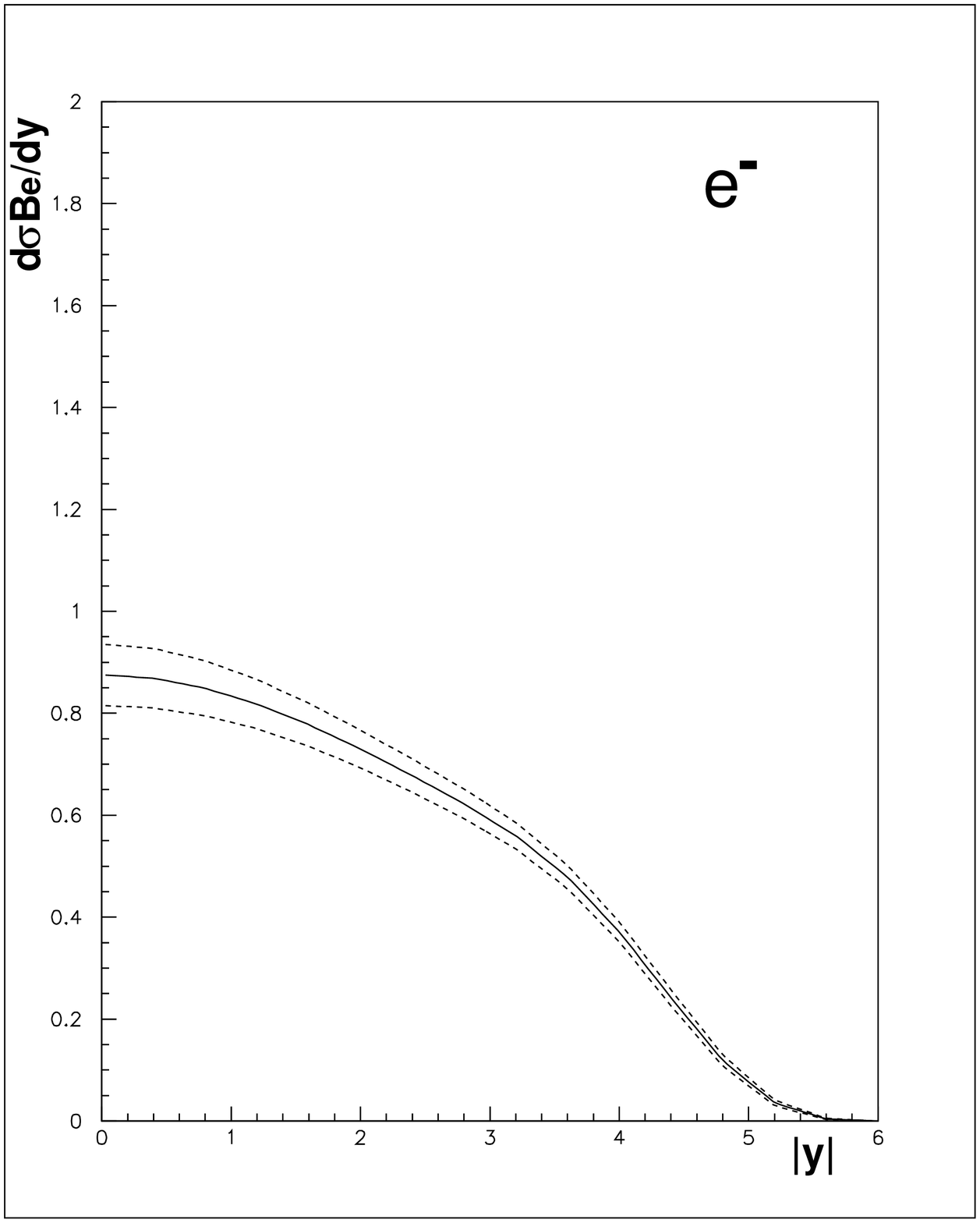,width=0.3\textwidth,height=4.5cm}
\epsfig{figure=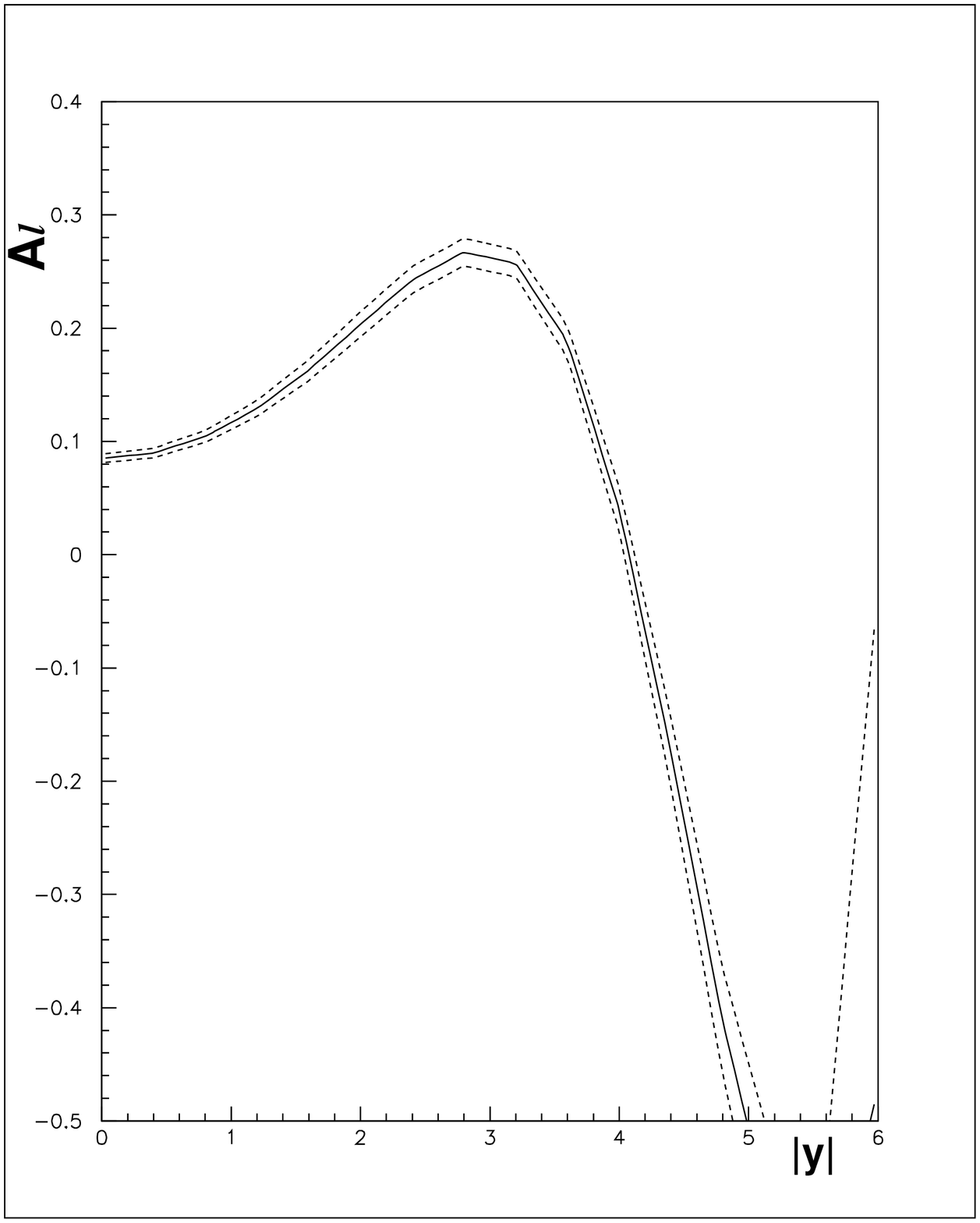,width=0.3\textwidth,height=4.5cm}
}
\centerline{
\epsfig{figure=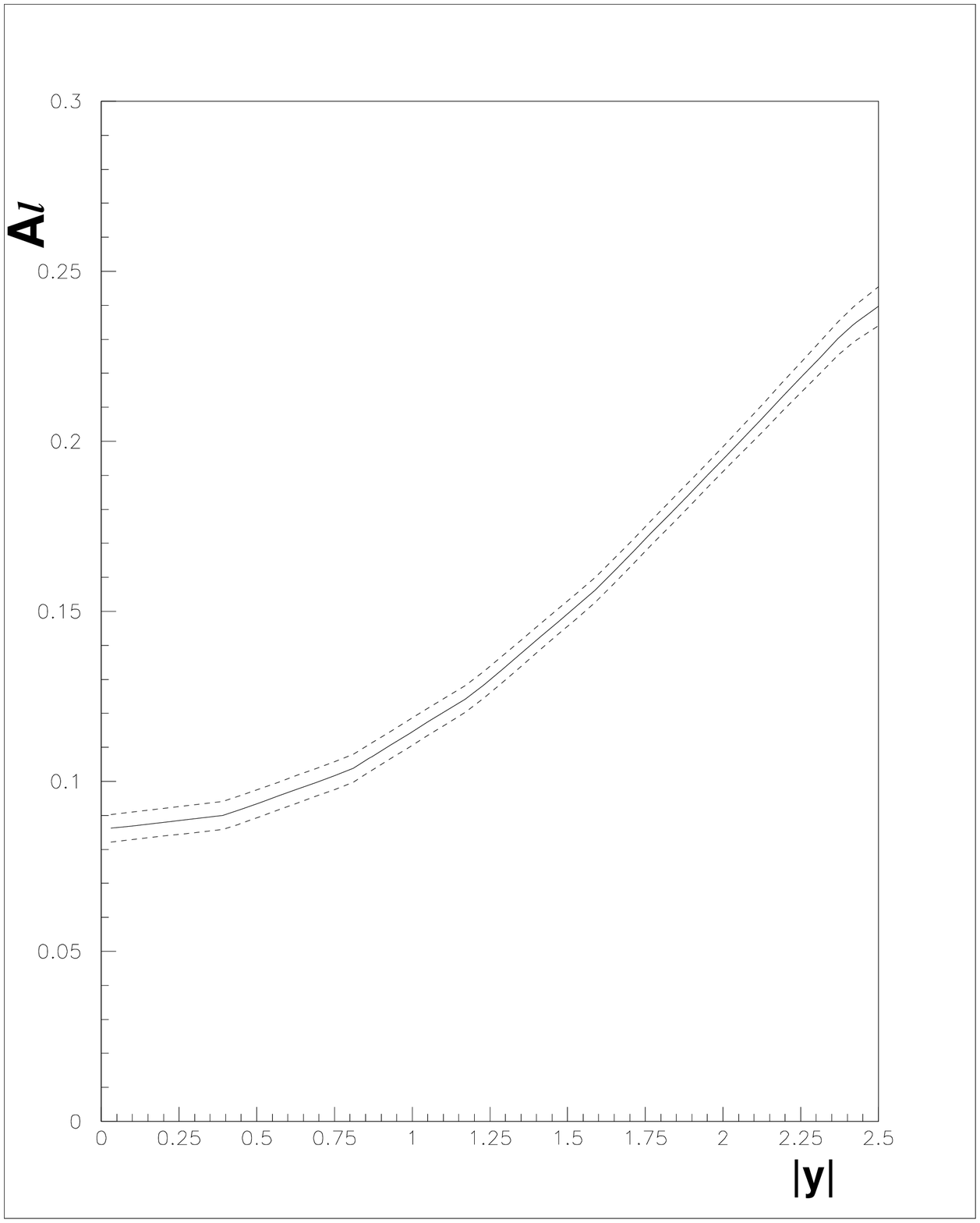,width=0.3\textwidth,height=4.5cm}
\epsfig{figure=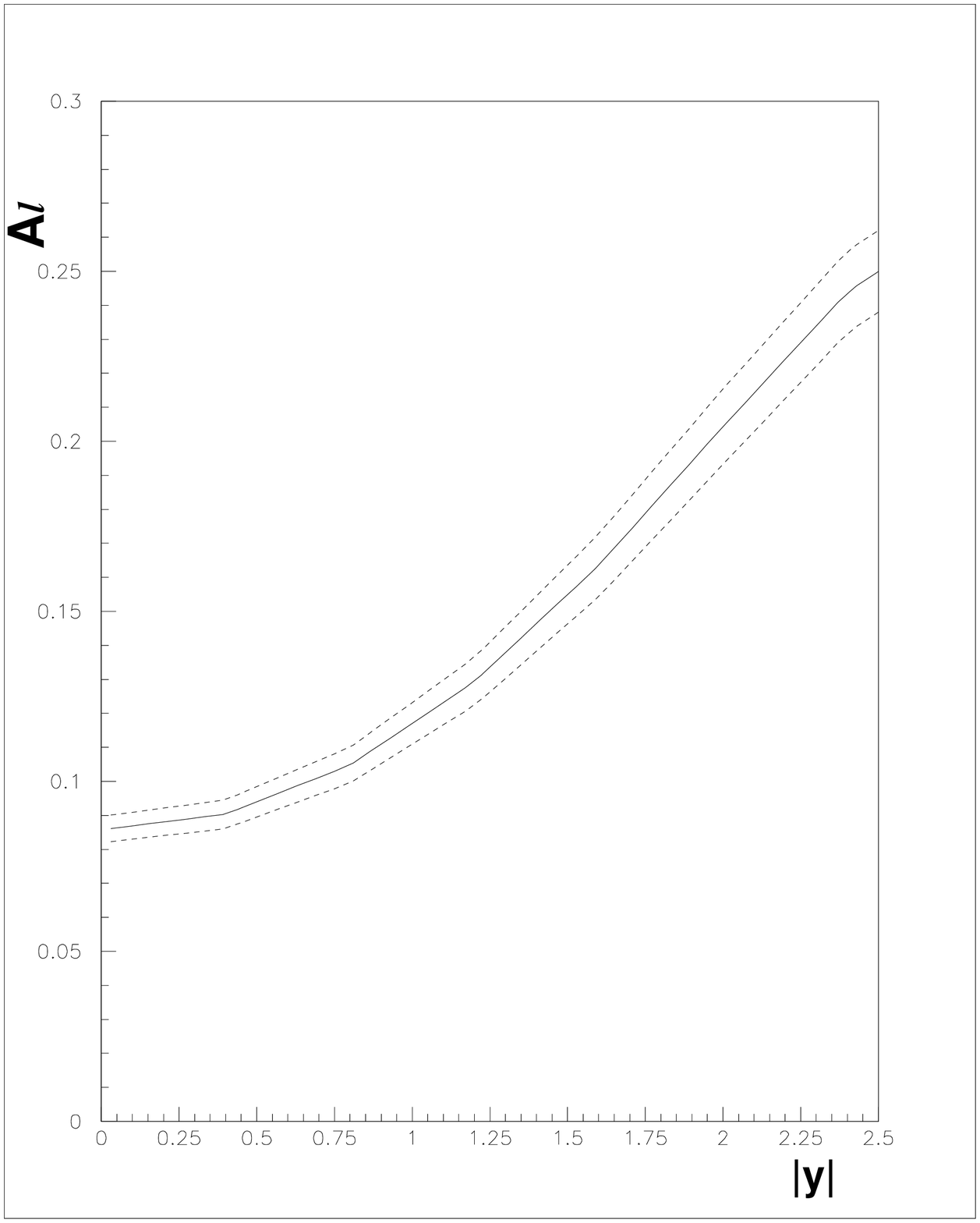,width=0.3\textwidth,height=4.5cm}
\epsfig{figure=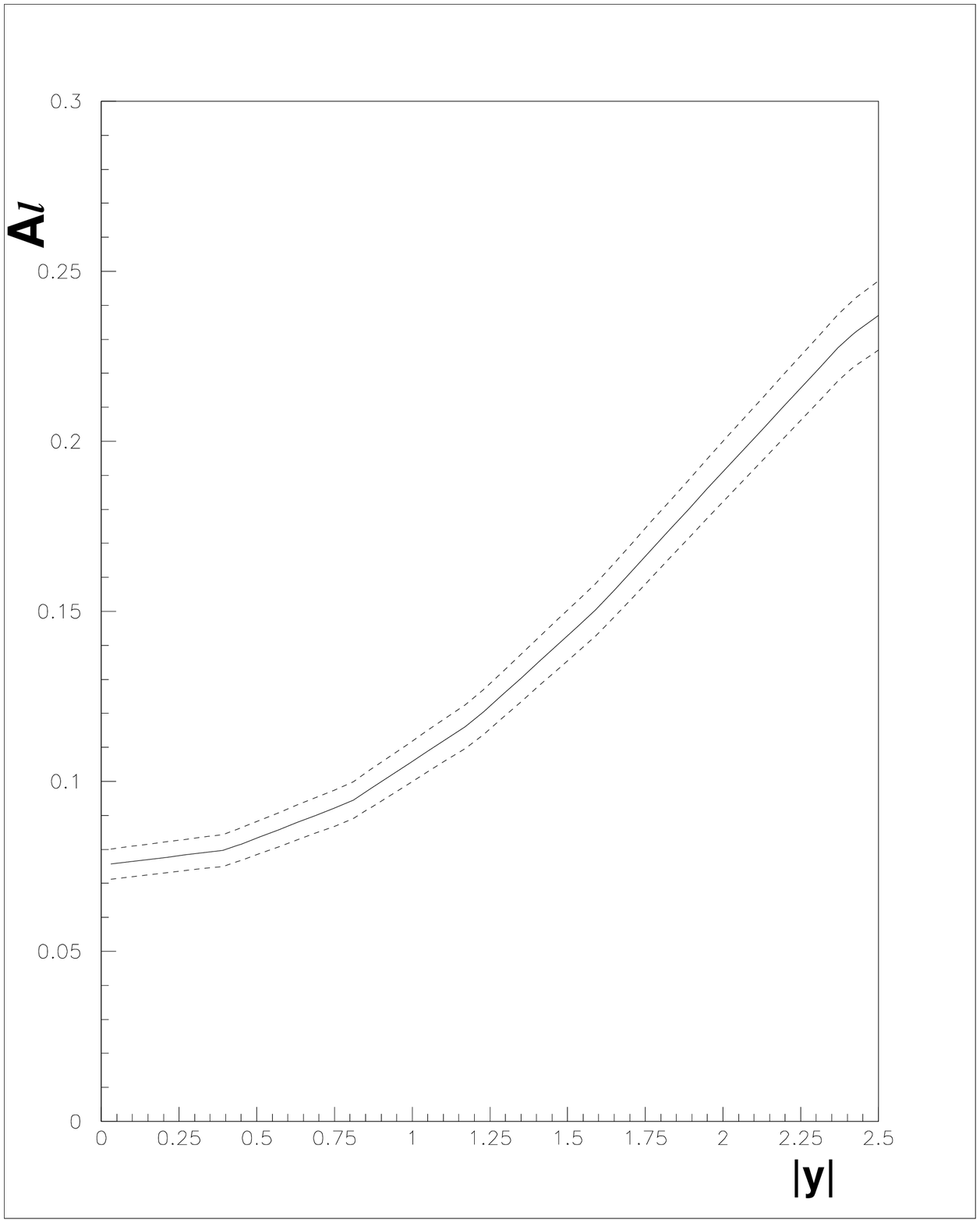,width=0.3\textwidth,height=4.5cm}
}
\caption {Top row: lepton spectra from the CTEQ6.1 PDFs; left plot, 
decay $e^+$ rapidity spectrum; middle plot, decay $e^-$ rapidity spectrum; 
right plot, lepton asymmetry, $A_l$. Bottom Row: the lepton asymmetry, 
$A_l$, from different PDF analyses; left plot ZEUS-S; 
middle plot CTEQ6.1; right plot MRST01}
\label{fig:leptons}
\end{figure}

To assess if LHC measurements will actually be discriminating we must first 
account for the fact that
$W$'s decay, and are most easily detected from their leptonic decays. 
Thus we actually measure the decay lepton 
rapidity spectra rather than the $W$ rapidity spectra. The upper half of  
Fig.~\ref{fig:leptons} shows these rapidity spectra for positive and 
negative leptons from $W^+$ and $W^-$ decay together with
 the lepton asymmetry, 
\[A_l = (l^+ - l^-)/(l^+ + l^-)\] for the CTEQ6.1 PDFs. 
A cut of, $p_{t} > 25$GeV, 
has been applied on the decay lepton, since it will not be possible to 
identify leptons with small $p_{t}$. 
A particular lepton rapidity can be fed from a range 
of $W$ rapidities so that the contributions of partons at different $x$ 
values are smeared out 
in the lepton spectra. Nevertheless the broad features 
of the $W$ spectra and the 
sensitivity to the gluon parameters are reflected in the lepton spectra,
resulting in a similar estimate ($\sim 8\%$) of PDF uncertainty  
at central rapidity for the CTEQ6.1 PDFs. 
The lepton asymmetry shows the change of sign at large $y$ which is 
characteristic of the $V-A$ 
structure of the lepton decay. The cancellation of the 
uncertainties due to the gluon PDF is not so 
perfect in the lepton asymmetry as in the $W$ asymmetry. Even so, in the 
measurable rapidity range, the PDF uncertainty in the asymmetry is smaller 
than in the lepton spectra, being $\sim 5\%$, for the 
CTEQ6.1 PDFs. The $Z$ to $W$ ratio $A_{ZW}$ has 
also been recalculated as a $Z$ to leptons ratio, \[A_{Zl} = Z/(l^+ +l^-)\] 
illustrated in Fig.~\ref{fig:awzw}. Just as for $A_{ZW}$, the overall 
uncertainty in $A_{Zl}$ is very small ($\sim 0.5\%$) for CTEQ6.1 PDFs.

It is again necessary to consider the difference between different 
PDF analyses for the predictions of the lepton spectra, $A_{Zl}$ and $A_l$. 
For the lepton spectra, the spread in the predictions of the 
different PDF analyses of MRST01, CTEQ6.1 and ZEUS-S is 
comparable to the uncertainty estimated by the individual analyses, just as 
for the $W$ spectra, and this is shown later in Fig.~\ref{fig:gendet}.  
Just as for $A_{ZW}$, there
are greater differences in the predictions for $A_{Zl}$ between PDF analyses 
than within any PDF analysis, but these differences remain within $\sim 5\%$
preserving this quantity as an SM benchmark measurement. Thus our 
previous estimate of the usefulness of these processes as luminosity monitors
and SM benchmarks
survives the reality check of the fact that we will measure the leptons, not
the $W$'s.

The significant differences which we noticed between the 
predictions of the different PDF analyses for $A_W$, remain in the predictions
for $A_l$. The lower half of Fig.~\ref{fig:leptons} 
compares these predictions for the ZEUS-S PDFs with those of 
the CTEQ6.1 PDFs and the MRST01 PDFs, in the measurable rapidity range.
The discrepancy of $\sim 25\%$ which was found in $A_W$ has been somewhat 
diluted to $\sim 15\%$ in $A_l$, but this should still be large enough for LHC 
measurements to discriminate, and hence to give information on the low-$x$
valence distributions. 

\section{HOW WELL CAN WE ACTUALLY MEASURE $W$ SPECTRA AT THE LHC?}
\label{sec:lowx;amcs_reality}

The remainder of this contribution will be concerned with the question: 
how accurately can we  
measure the lepton rapididty spectra and can we use the early LHC data 
to improve on the current level of uncertainty?

We have simulated one million signal, $W \rightarrow e \nu_e$, events for 
each of the PDF sets CTEQ6.1, MRST2001 and ZEUS-S using HERWIG (6.505). 
For each of these PDF sets the eigenvector error PDF sets have been simulated 
by PDF reweighting and k-factors have been applied to approximate an NLO generation. A study has been made of the validity of both PDF reweighting and 
k-factor reweighting and this is reported in ref.~\cite{hep-ex/0509002}. The 
conclusion is that PDF reweighting is valid for reweighting the rapidity 
spectra when the PDF sets are broadly similar, as they are within any one PDF 
analysis. The k-factor reweighting to simulate NLO is also valid for the 
rapidity spectra for which it was designed.

\begin{figure}[tbp] 
\vspace{-1.0cm}
\centerline{
\epsfig{figure=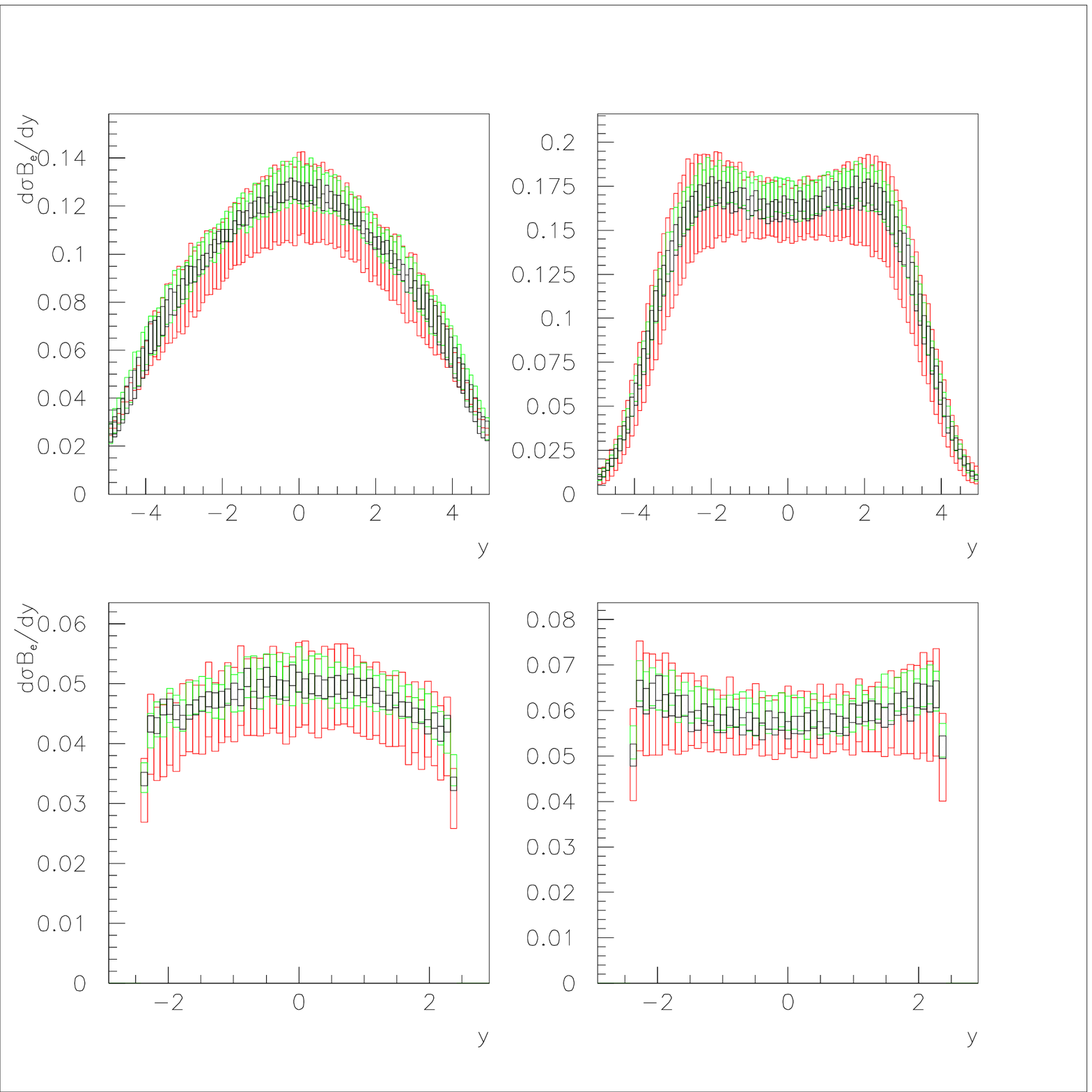,width=0.66\textwidth,height=9cm }
\epsfig{figure=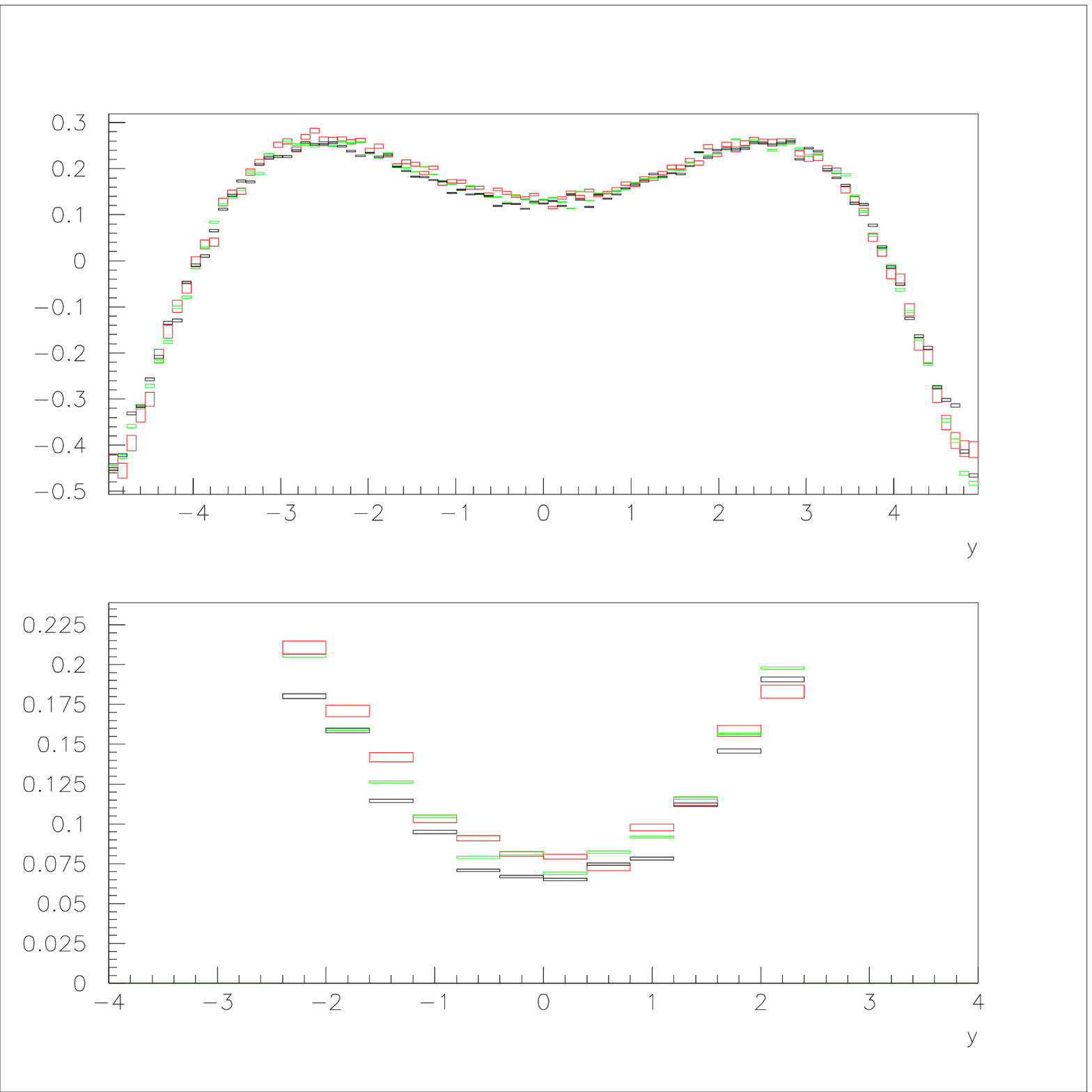,width=0.33\textwidth,height=9cm}
}
\caption {Top row: $e^-$, $e^+$ and $A_e$ rapidity spectra for the lepton from the $W$ decay, 
generated using HERWIG + k factors and CTE6.1 (red),
ZEUS-S (green) and MRST2001 (black) PDF sets with full uncertainties. Bottom row: the same spectra after passing 
through the ATLFAST detector simulation and selection cuts.}
\label{fig:gendet}
\end{figure}

The top part of Fig.~\ref{fig:gendet}, shows the $e^{\pm}$ and $A_l$ 
spectra at the generator level, for all of the PDF sets sumperimposed. As mentioned before, it is 
clear that the lepton spectra as predicted by the different PDF analyses are 
compatible, within the PDF uncertainties of the analyses.
The events are then passed through the ATLFAST fast simulation of the ATLAS 
detector. This applies loose kinematic cuts: 
$|\eta| < 2.5$, $p_{te} > 5 GeV$, and electron isolation criteria. 
It also smears the 4-momenta of the 
leptons to mimic momentum dependent detector resolution. 
We then apply further cuts designed to eliminate the background preferentially. These criteria are:
\begin{itemize}
\item pseudorapidity, $|\eta| <2.4$, to avoid bias at the edge of the measurable rapidity range
\item  $p_{te} > 25 GeV$, high $p_t$ is necessary for efficient electron identification
\item  missing $E_t > 25$ GeV, the $\nu_e$ in a signal event will have a correspondingly large missing $E_t$
\item  no reconstructed jets in the event with $p_t > 30 GeV$, to discriminate against QCD background 
\item  recoil on the transverse plane $p_t^{recoil} < 20GeV$, to discriminate against QCD background
\end{itemize}
These cuts ensure that background from the processes: 
$W \rightarrow \tau \nu_\tau$; $Z \rightarrow \tau^+ \tau^-$; and $Z \rightarrow e^+ e^-$, is negligible ($\leqsim 1\%$)~\cite{hep-ex/0509002}. Furthermore, 
a study of charge misidentification has established that the lepton asymmetry 
will need only very small corrections ($\leqsim 0.5\%$), 
within the measurable rapidity range~\cite{hep-ex/0509002}.

The lower half of
Fig.~\ref{fig:gendet}, shows the $e^{\pm}$ and $A_l$ spectra at the detector level after application 
of these cuts, for all of the PDF sets superimposed. 
The level of precision of each PDF set, seen in the analytic calculations of Fig.~\ref{fig:mrstcteq}, appears 
somewhat degraded  at detector level, so that a net level of PDF 
uncertainty in the lepton spectra of $\sim 10\%$ is expected at central 
rapidity. Thus the usefulness of these processes as a luminosity monitor
is somewhat compromised if a measurement to better than $10\%$ is required.

The anticipated cancellation of PDF 
uncertainties in the asymmetry spectrum is observed, within each PDF set, 
such that the uncertainties predicted by each PDF set are $\sim 5\%$, but the 
spread between the MRST and CTEQ/ZEUS-S PDF sets is as large as $\sim 15\%$.
Thus measurements which are accurate to about $\sim 5\%$ could provide useful
information on the valence distributions at low $x$.

\section{USING LHC DATA TO IMPROVE PRECISION ON PDFs}
\label{sec:lowx;amcs_improve}

We now consider the possibility of improving on the current level of PDF 
uncertainty by using LHC data itself.
 The high cross-sections for $W$ prodution at 
the LHC ensure that it will be the experimental systematic errors, rather than the statistical errors, which 
are determining. Our experience with the detector simulation leads us to believe that a systematic precision of $\sim 5\%$ could be achievable. 
We have optimistically imposed a random  $4\%$
scatter on our samples of one million $W$ events, 
generated using different PDFs, in order to 
investigate if measurements at this level of precision will improve PDF 
uncertainties at central rapidity significantly, if they 
are input to a global PDF fit. 

\begin{figure}[tbp] 
\vspace{-1.5cm}
\centerline{
\epsfig{figure=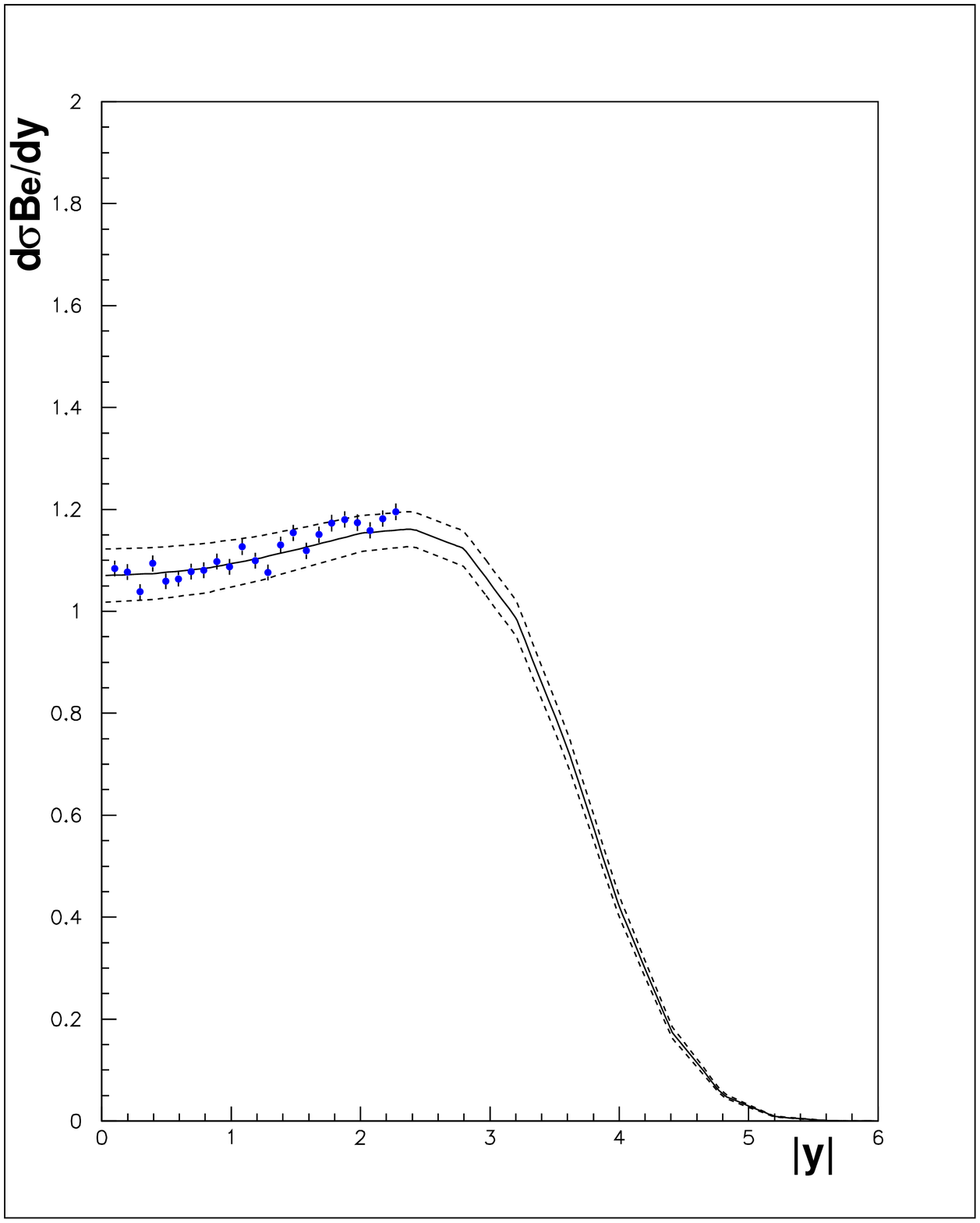,width=0.3\textwidth,height=4cm}
\epsfig{figure=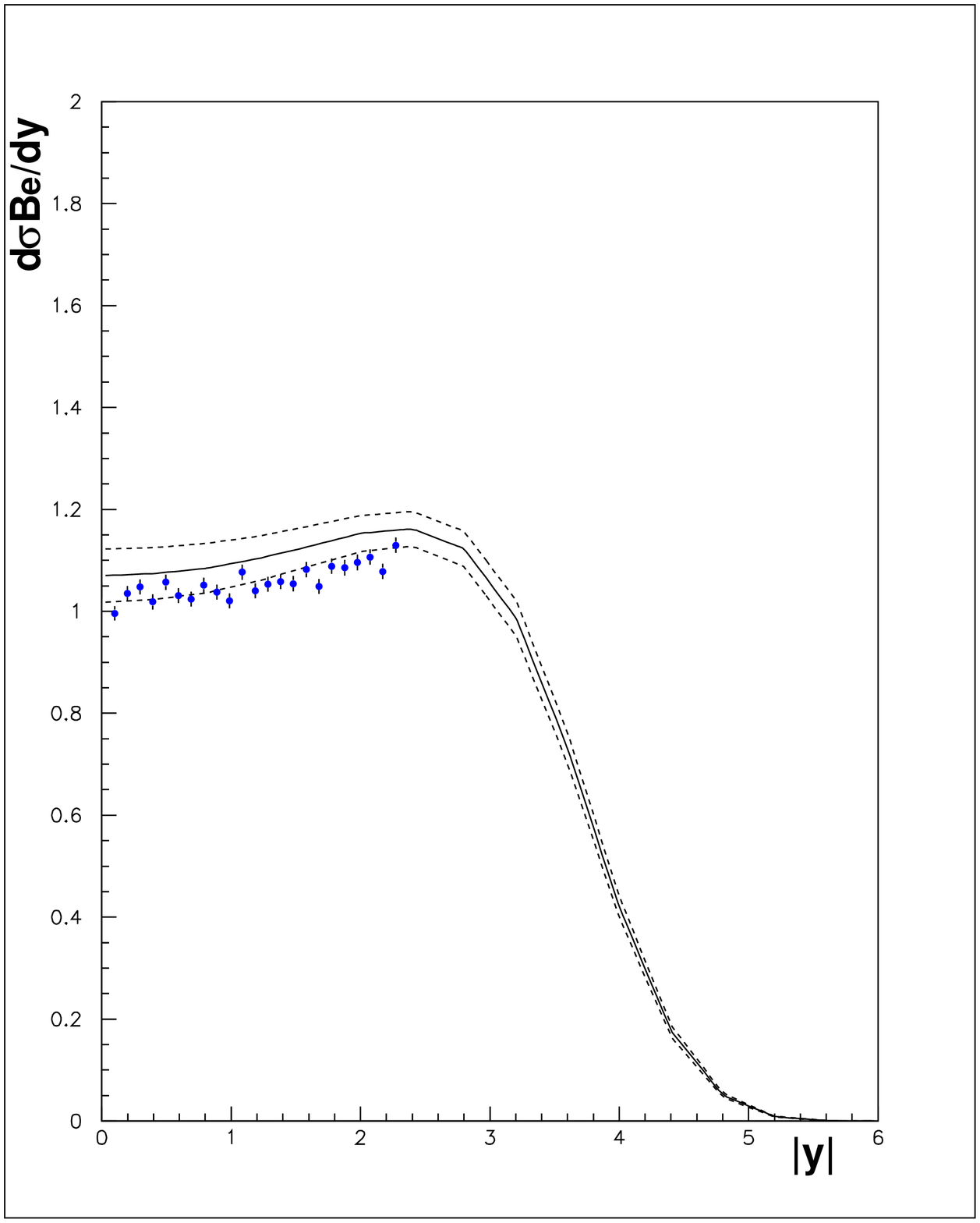,width=0.3\textwidth,height=4cm}
\epsfig{figure=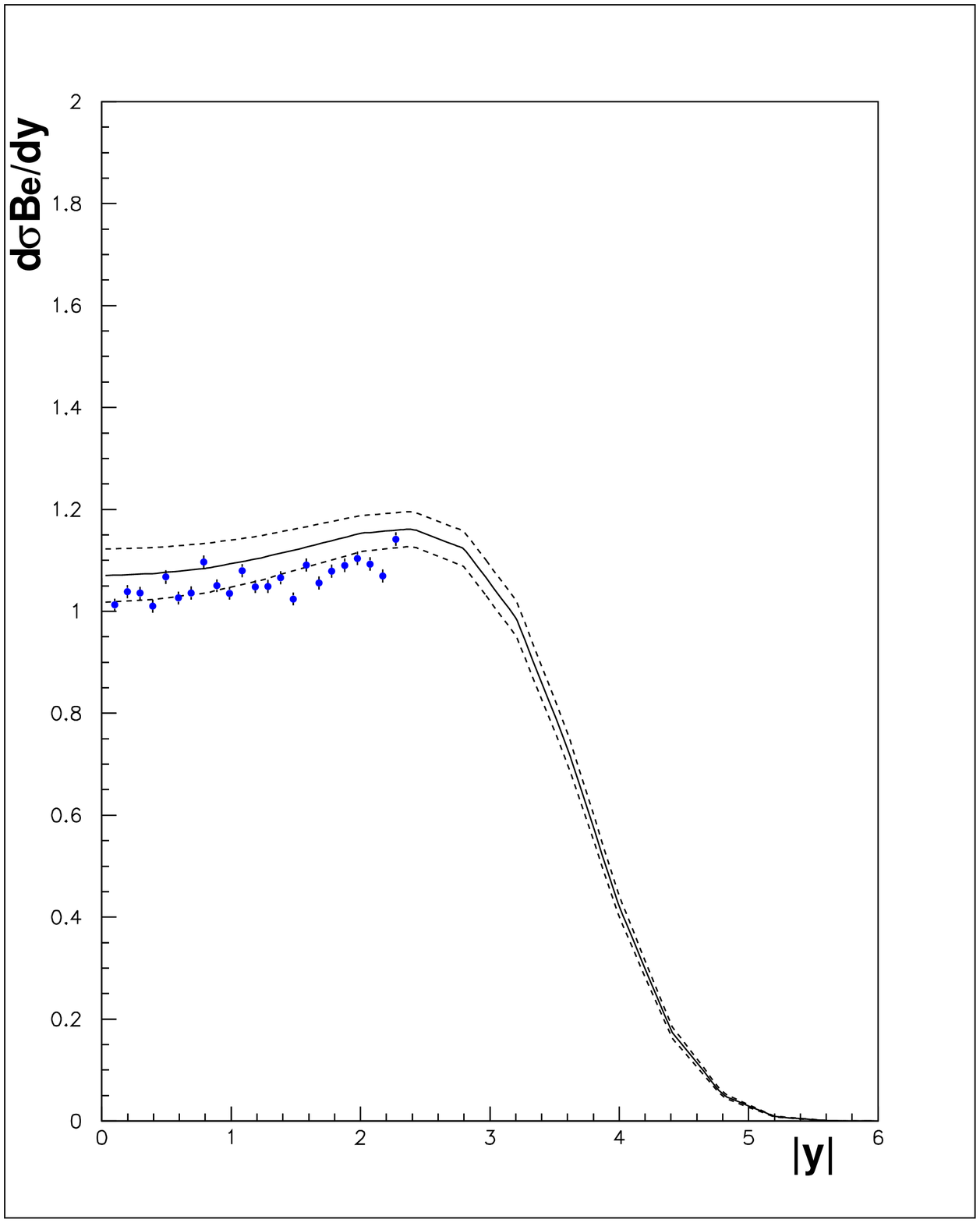,width=0.3\textwidth,height=4cm}
}
\centerline{
\epsfig{figure=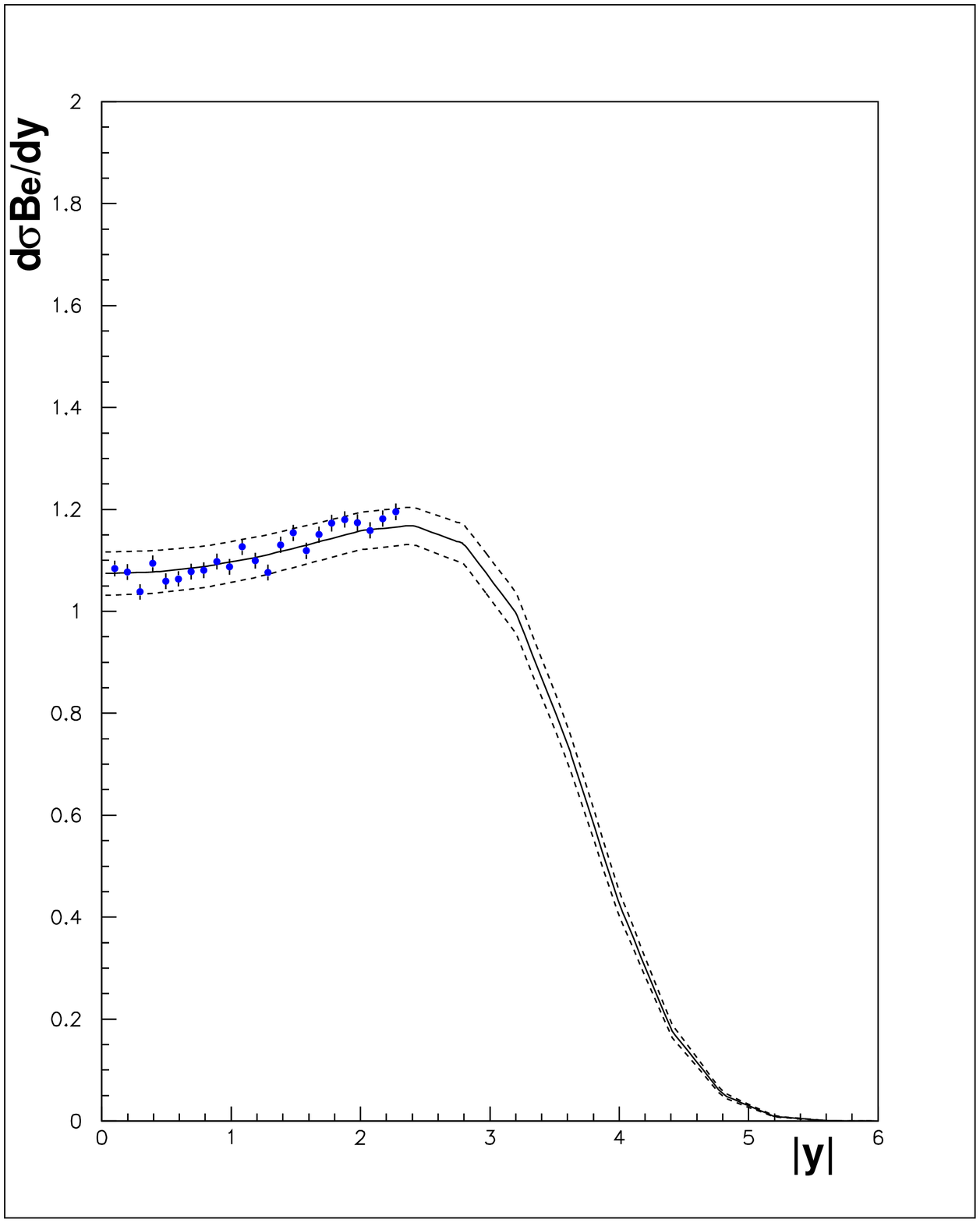,width=0.3\textwidth,height=4cm}
\epsfig{figure=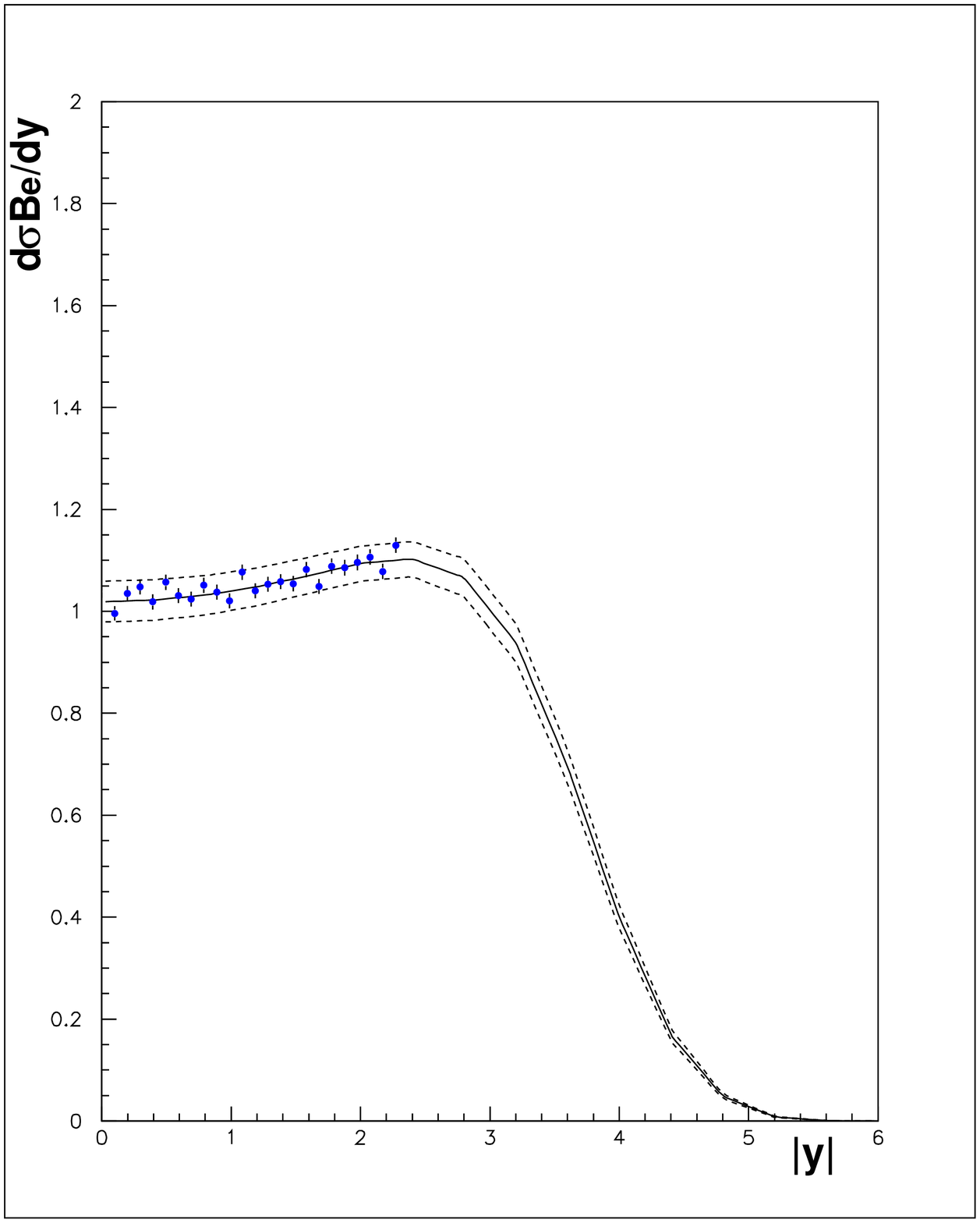,width=0.3\textwidth,height=4cm}
\epsfig{figure=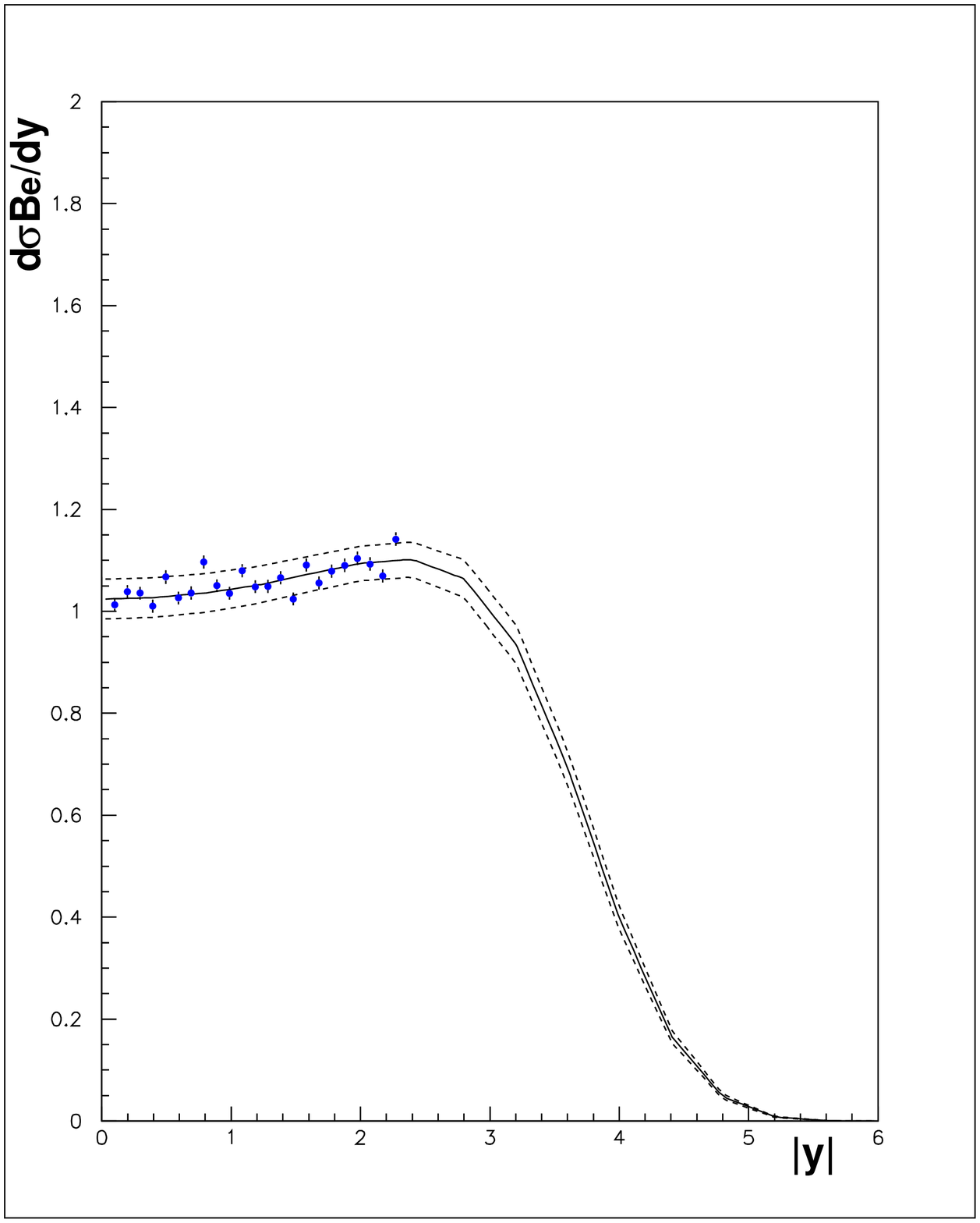,width=0.3\textwidth,height=4cm}
}
\caption {Top row: $e^+$ rapidity spectra generated from: left plot, ZEUS-S 
PDFS; middle plot, CTEQ6.1 PDFs; right plot, CTEQ6.1 PDFs which have been 
passed through the ATLFAST detector simulation and corrected back to generator
level using ZEUS-S PDFs; compared to the analytic prediction
using ZEUS-S PDFs. Bottom row: the same lepton rapidity spectra as above 
compared to the analytic 
prediction AFTER including these lepton pseudo-data in the ZEUS-S PDF fit.}
\label{fig:zeusfit}
\end{figure}

The upper left hand plot of Fig.~\ref{fig:zeusfit} shows the $e^+$ rapidity 
spectra for events generated from the ZEUS-S PDFs compared to the analytic 
predictions for these same
 ZEUS-S PDFs. The lower left hand plot illustrates the result if 
these events are then 
included in the ZEUS-S PDF fit (together with the $e^-$ spectra which are 
not illustrated). The size of the PDF uncertainties, at $y=0$, 
decreases from $6\%$ to $4.5\%$.  
The largest improvement is in the PDF parameter $\lambda_g$ controlling the 
low-x gluon at the input scale, $Q^2_0$: $xg(x) \sim x^{\lambda_g}$ at 
low-$x$, $\lambda_g = -0.199 \pm 0.046$, 
before the input of the LHC pseudo-data, compared to, 
$\lambda_g = -0.196 \pm 0.029$, after input. 
Note that whereas the relative normalisations of the $e^+$ and $e^-$ spectra 
are set by the PDFs, 
the absolute normalisation of the data is free in the fit so that 
no assumptions are made on our ability to measure luminosity.
Secondly, we repeat this procedure for events generated using the CTEQ6.1 PDFs.
This is illustrated in the middle section of Fig.~\ref{fig:zeusfit}. 
Before they are input to the fit, the cross-section for these events is on the 
lower edge of the uncertainty band of the ZEUS-S predictions 
(upper middle plot). If these events are then input to the fit the 
central value shifts and the uncertainty decreases (lower middle plot). 
The value of of the parameter $\lambda_g$ becomes, 
$\lambda_g = -0.189 \pm 0.029$, after input of these pseudo-data.
Finally, to simulate the situation which really faces experimentalists, 
we generate events with CTEQ6.1, 
and pass them through the ATLFAST detector simulation and cuts. We then 
correct back from detector level 
to generator level using a different PDF set- in this cases the ZEUS-S PDFs- 
since in practice we will not know 
the true PDFs. The upper right hand plot of Fig.~\ref{fig:zeusfit} shows 
that the resulting corrected data look 
pleasingly like CTEQ6.1, but they are more smeared. When these data are input 
to the PDF fit the central values shift and 
errors decrease (lower right plot) just as for the perfect CTEQ6.1 
pseudodata. The value of $\lambda_g$ becomes,
 $\lambda = -0.181 \pm 0.030$, after input of these pseudodata. 
Thus we see that the bias introduced by the 
correction procedure from detector to generator level is small compared to 
the PDF uncertainty, and that measurements at the $\sim 4\%$ level should be 
able to improve the level of uncertainty of the PDF predictions.

\section*{CONCLUSIONS}
\label{sec:lowx;amcs_conc}

We have investigated the PDF uncertainty on the predictions for $W$ and $Z$ production at the LHC, using the electron decay channel for the $W$s and taking 
into account realistic expectations for measurement accuracy and the cuts on data which will be needed to 
identify signal events from background processes. We conclude that, at the present level of PDF uncertainty, the 
decay lepton spectra can be used as a luminosity monitor but it is only good 
to $\sim 10\%$. However, 
we have also investigated the measurement accuracy 
necessary for early measurements of these decay lepton spectra to be useful in further constraining the 
PDFs. A systematic measurement error of $\sim 4\%$ could provide 
useful extra constraints.

The ratio of $Z$ to $W^+ + W^-$ production (measured via the lepton spectra) can provide an SM measurement
which is relatively insensitive to PDF uncertainties. 
By contrast a measurement of the lepton asymmetry can provide the 
first measurements of the valence difference $u_v - d_v$ at small $x$.

We now return to the caveat made in the introduction: 
the current study has been performed using 
standard PDF sets which are extracted using NLO QCD in the 
DGLAP formalism. The extension to NNLO is 
straightforward, giving small corrections $\sim 1\%$. PDF analyses at NNLO 
including full accounting of the PDF 
uncertainties are not extensively available yet, so this small correction 
has not been pursued here. However, there may be much larger  
uncertainties in the theoretical calculations because the kinematic region 
involves  low-$x$.  
The MRST group recently produced a PDF set, MRST03, which does not include 
any data for $x < 5\times 10^{-3}$, in order to avoid 
the inappropriate use of the DGLAP formalism at small-$x$. 
Thus the  MRST03 PDF set should only be used for $x > 5\times 10^{-3}$. 
What is needed is an alternative theoretical formalism
for smaller $x$, as suggested by R. Ball in these proceedings. It is clear 
that the use of this formalism would bring greater changes than the small 
corrections involved in going to NNLO. There may even  a need for  
more radical extensions of the theory at low-$x$ due to high density effects.

\begin{figure}[tbp] 
\centerline{
\epsfig{figure=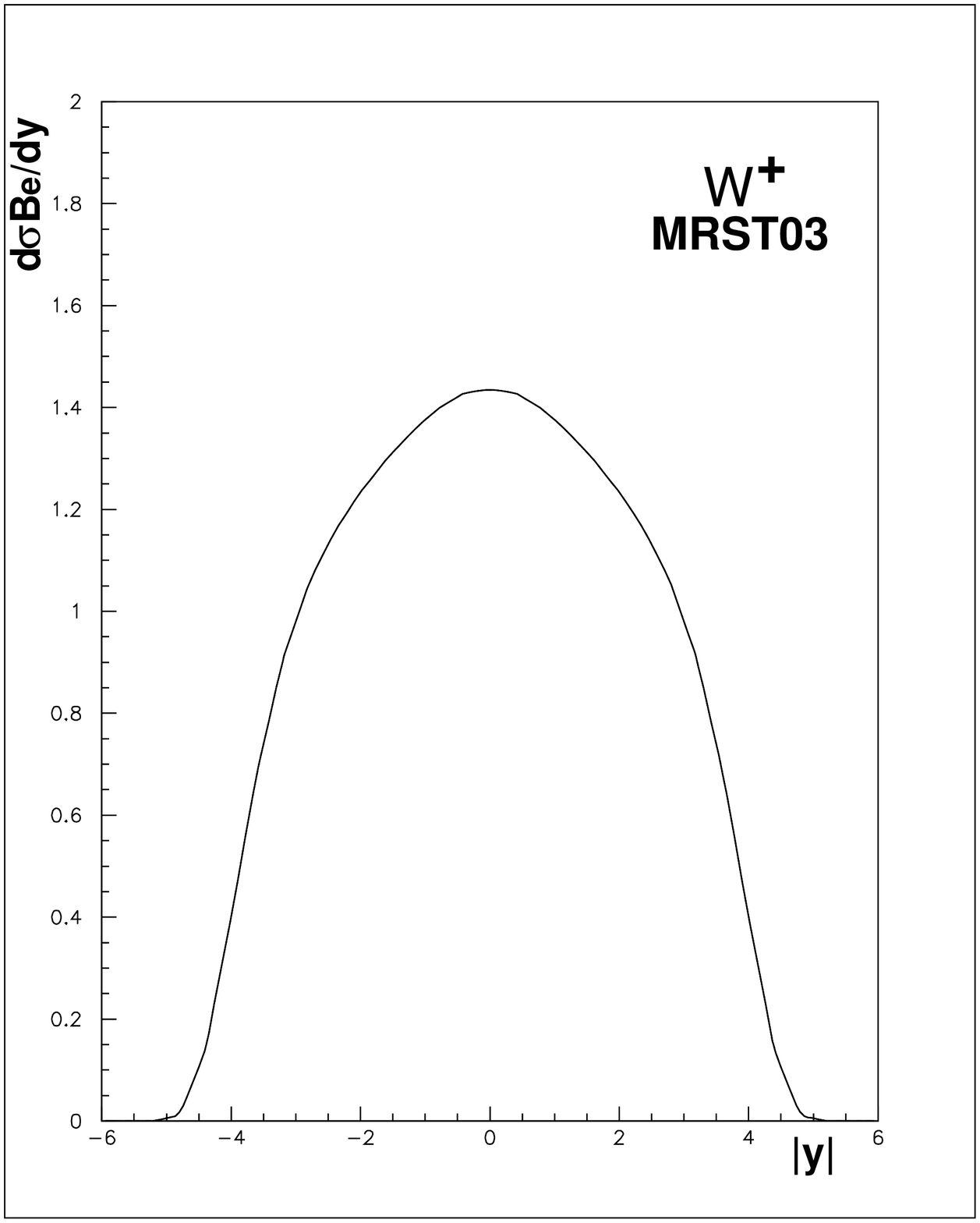,width=0.3\textwidth,height=4cm}
\epsfig{figure=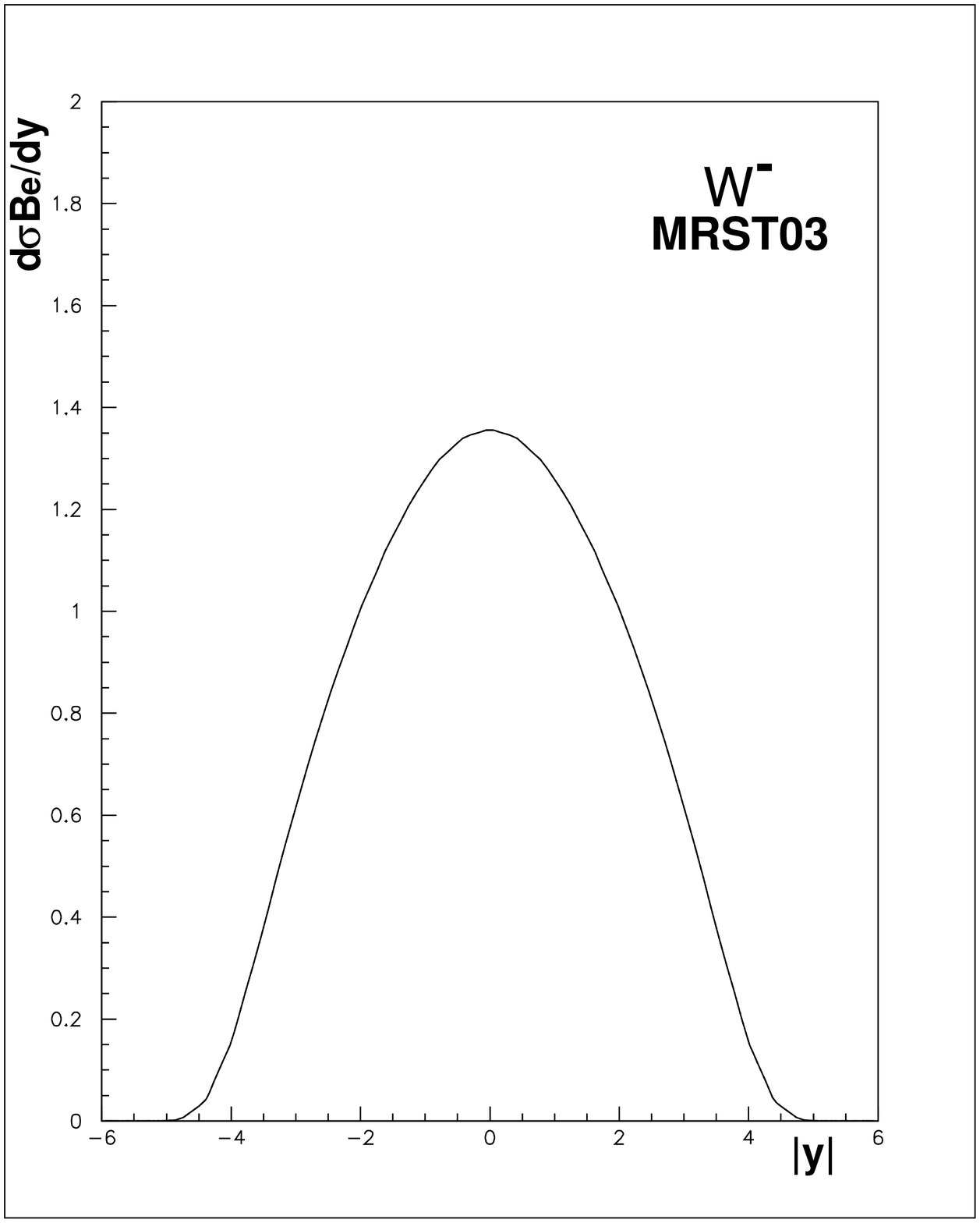,width=0.3\textwidth,height=4cm}
\epsfig{figure=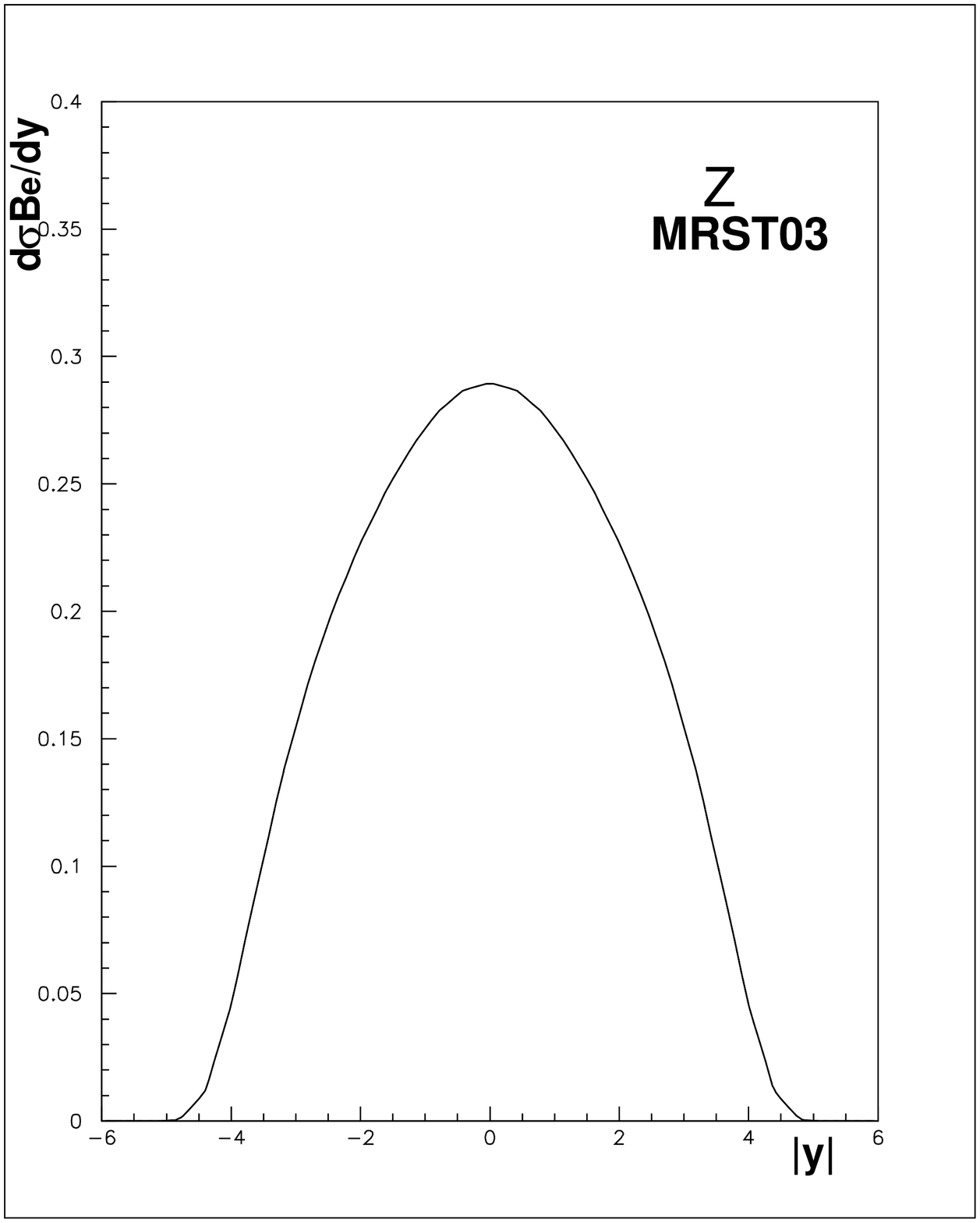,width=0.3\textwidth,height=4cm} 
}
\caption {LHC $W^+,W^-,Z$ rapidity distributions for the MRST03 PDFs: left plot $W^+$; middle plot $W^-$; 
right plot $Z$}
\label{fig:mrst03pred}
\end{figure}

The MRST03 PDF set may be used as a toy PDF set, 
to illustrate the effect of using 
very different PDF sets on our predictions. 
A comparison of Fig.~\ref{fig:mrst03pred} with 
Fig.~\ref{fig:WZrapFTZS13} or Fig.~\ref{fig:mrstcteq} shows how different 
the analytic predictions are from the conventional ones, and thus illustrates 
where we might  expect to see differences due 
to the need for an alternative formalism at small-$x$. Whereas these results 
may seem far fetched we should remind ourselves that moving into a different 
kinematic regime can provide suprises- as it did with the HERA data itself!  

\bibliography{coopersarkar}

\end{document}